\documentclass[10pt]{iopart}
\usepackage{hyperref}
\usepackage{tikz}
\usepackage{subfigure}
\usetikzlibrary{arrows, decorations.markings}
\usepackage{rotating}
\usepackage{xcolor}
\usepackage{colortbl} 
\usepackage{graphicx}

\usepackage{capt-of}
\usepackage{booktabs}
\usepackage{varwidth}
\newsavebox\tmpbox

\expandafter\let\csname equation*\endcsname\relax
\expandafter\let\csname endequation*\endcsname\relax
\usepackage[version=4]{mhchem} 

\usepackage{cleveref}
\usepackage[]{units}

\usepackage{booktabs}      
\usepackage{multirow}

\usepackage{enumitem}

\newcommand{\mub}{\,\mu_\text{B}}

\newcommand{\ttwog}{t$_{2\mathrm{g}}$ }

\definecolor{orange}{HTML}{FDE9D7}
\definecolor{grey}{HTML}{F4F4F4}

\newcommand{\white}[1]{\textcolor{white}{#1}}
\newcommand{\red}[1]{\textcolor{black}{#1}}

\begin{document}
\title{Magnetic structures and electronic properties of cubic-pyrochlore ruthenates from first principles}
\author{M.-T.\ Huebsch$^{1,2}$, Y.\ Nomura$^1$, S.\ Sakai$^1$ and R.\ Arita$^{1,3}$}
\address{$^1$Center for Emergent Matter Science, RIKEN, Wako, Saitama 351-0198, Japan}
\address{$^2$VASP Software GmbH, Sensengasse 8/17, A-1090 Vienna, Austria}
\address{$^3$University of Tokyo, 7-3-1 Hongo, Bunkyo-ku, Tokyo 113-8656, Japan}
\date{\today}

\ead{marie-therese.huebsch@vasp.at}

\begin{abstract}
  The magnetic ground states of $R_2$Ru$_2$O$_7$ and $A_2$Ru$_2$O$_7$ with $R=$ Pr, Gd, Ho, and Er, as well as $A=$ Ca, Cd are predicted devising a combination of the cluster-multipole (CMP) theory and spin-density-functional theory (SDFT). The strong electronic correlation effects are estimated by the constrained-random-phase approximation (cRPA) and taken into account within the dynamical-mean-field theory (DMFT). 
  The target compounds feature $d$-orbital magnetism on Ru$^{4+}$ and Ru$^{5+}$ ions for $R$ and $A$, respectively, as well as $f$-orbital magnetism on the $R$ site, which leads to an intriguing interplay of magnetic interactions in a strongly correlated system. 
  We find CMP+SDFT is capable of describing the magnetic ground states in these compounds. 
  The cRPA captures a difference in the screening strength between $R_2$Ru$_2$O$_7$ and $A_2$Ru$_2$O$_7$ compounds, 
  which leads to a qualitative and quantitative understanding of the electronic properties within DMFT.
\end{abstract}


\submitto{\JPCM}

\maketitle
\ioptwocol

\section{Introduction} \label{sec:Introduction}

Cubic-pyrochlore ruthenates \cite{gardner2010magnetic,rau2019frustrated} are subject to magnetic frustration, strong electronic correlation, and in some cases a considerable amount of spin-orbit coupling (SOC).
Consequently, this family of materials, that has the chemical formula $(R_{1-x}A_x)_2$Ru$_2$O$_7$, displays a plethora of different phases and transitions, including the Mott-insulating state, bad metal state, spin-ice-like states and other noncollinear magnetism. The ground state strongly depends on the choice of cations $A^{2+}$ and $R^{3+}$. Yet, in addition to that, different phases can be reached by introducing hole doping $x$ \cite{ueda2020evolution,kaneko2021fully} or by applying pressure \cite{jiao2018effect}. Curiously, the experimentally observed tendencies with changing conditions are sometimes counter-intuitive, as described below. 
Here, we demonstrate that first-principles calculations can help elucidate the underlying phenomena of these counter-intuitive tendencies.

\begin{table*}
    \caption{Crystallographic positions for the space group $Fd\bar{3}m$ (No.\ 227) of the cubic-pyrochlore structure $(R_{1-x}A_x)_2$Ru$_2$O$_6$O$'$ with origin at $16c$. The parameter $x_a$ is the only compound-dependent parameter, apart from the lattice constant $a$. \red{For more details about the crystallographic information, we refer to Ref. \cite{gardner2010magnetic}.}
    \label{tab:1}
    }
    
    \begin{indented}
    \item[]\begin{tabular}{@{}llll}
    \br
    Atom              & Wyckoff position   & Site symmetry     & Coordinate \\ \mr 
    $A$/$R$           & 16$d$              & $\bar{3}m$($D_{3d}$)  & $\frac{1}{2}$, $\frac{1}{2}$, $\frac{1}{2}$ \\ 
    Ru                & 16$c$              & $\bar{3}m$ ($D_{3d}$) & $0$, $0$, $0$   \\ 
    O                 & 48$f$              & $mm$ ($C_{2v}$) & $x_a$, $\frac{1}{8}$,$\frac{1}{8}$   \\ 
    O$'$            & 8$b$               & $\bar{4}3m$ ($T_{d}$) & $\frac{3}{8}$, $\frac{3}{8}$,$\frac{3}{8}$   \\ \br
    \end{tabular}
    \end{indented}
\end{table*}

\begin{figure}
  \includegraphics[width=\columnwidth]{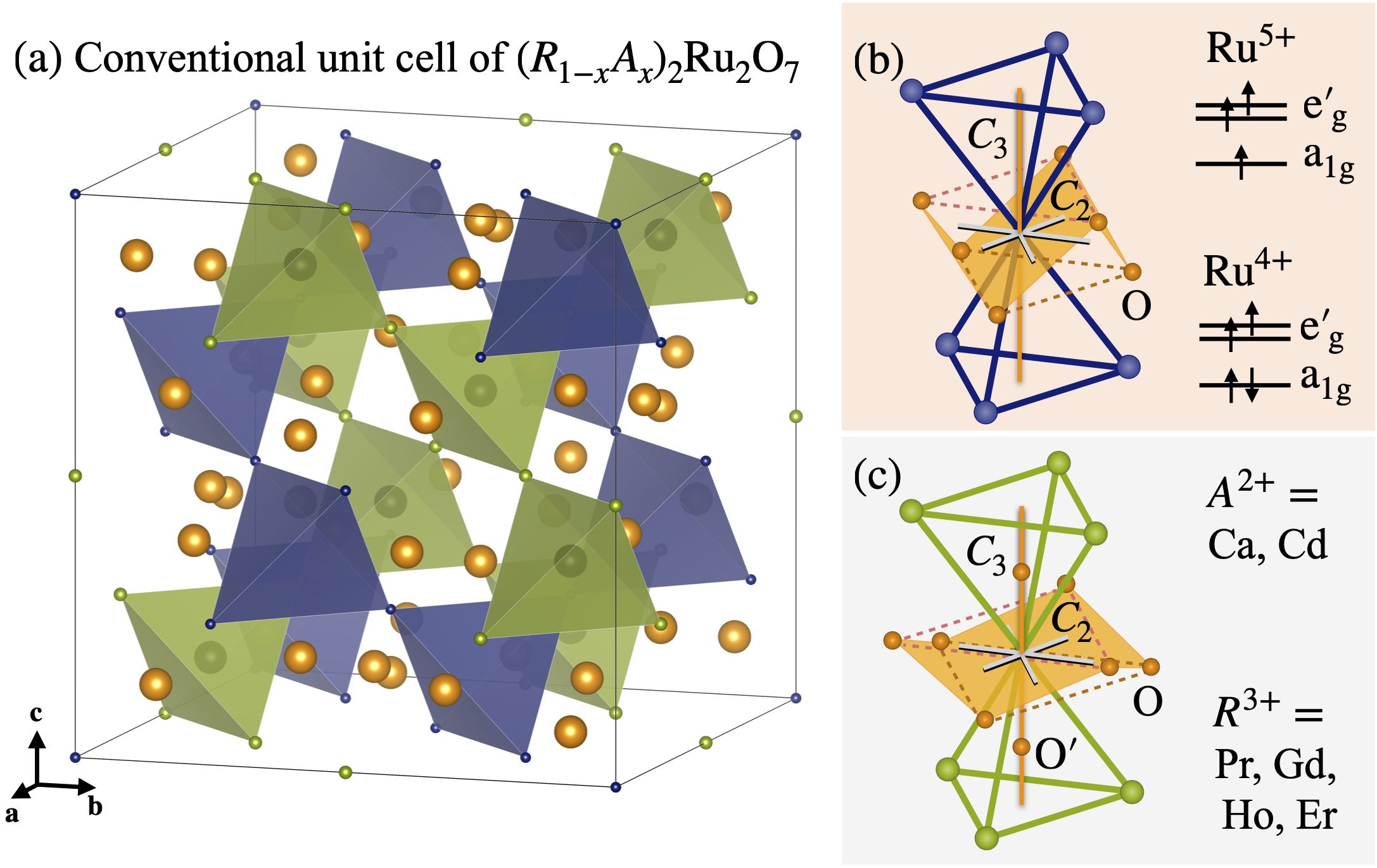}
  \caption{Crystal structure of $(R_{1-x}A_x)_2$Ru$_2$O$_6$O$'$. (a) Conventional unit cell showing the cubic-pyrochlore Ru ($R/A$) sublattice in blue (green) and O in orange. $\mathrm{O}'$ is located inside the green $R/A$ tetrahedra. (b) and (c) show the local environment of Ru and $R/A$, which both display $\bar{3}m$($D_{3d}$) site symmetry. The crystal-electric-field splitting and occupancy of the Ru-t$_{2\mathrm{g}}$ manifold into a lower a$_{1\mathrm{g}}$ and upper e$'_\mathrm{g}$ bands for Ru$^{5+}$ (Ru$^{4+}$) that occurs for the choice of $A$ ($R$) element is depicted at the top (bottom), in (b) and (c) respectively.
  \label{fig:1}
  }
\end{figure}

In $(R_{1-x}A_x)_2$Ru$_2$O$_7$, magnetic frustration arises from the corner sharing tetrahedra formed by both, $R$ and Ru sites, as shown in \Cref{fig:1} (a). The crystallographic details can be seen at a glance in \Cref{tab:1}. 
In particular, we will consider the cases with a magnetic rare-earth ion $R^{3+}=$ Pr, Gd, Ho and Er, and with a nonmagnetic cation $A^{2+}=$ Ca, and Cd in details. This choice covers a wide range of magnetic and electronic phases with different prevalent mechanisms.

For rare-earth ruthenates, $R_2$Ru$_2$O$_7$, the cubic-pyrochlore structure amounts to an interesting constellation of two magnetically coupled frustrated sublattices at 16$c$ and 16$d$, that are dominated by $d$-orbital and $f$-orbital magnetism, respectively. 
In contrast, in $A_2$Ru$_2$O$_7$ only one magnetic site, i.e., the Ru site, exists. Still, as discussed in \Cref{sec:Magnetic structure}, the $d$-orbital magnetism is fundamentally different in the limiting cases of $(R_{1-x}A_x)_2$Ru$_2$O$_7$ owing to an effective integer and half-integer spin state on the Ru site.

Capturing the subtleties of cubic-pyrochlore ruthenates on the experimental side is a challenge in itself, as indicated by the relatively recently successful synthesis, low transition temperatures and remaining uncertainties of the magnetic-structure measurements, c.f., \Cref{sec:Magnetic structure}. 
\red{A prominent example of the possible magnetic configurations is the spin-ice structure which is characterized by its magnetic entropy \cite{bramwell2001science,gardner2005spin}.}
Regarding the electronic properties, recent advances by Kaneko \emph{et al.\ }\cite{kaneko2021fully} made it possible to investigate fully filling-controlled (Ca$_{1-x}$Pr$_x)_2$Ru$_2$O$_7$, which shows a metal-to-insulator transition (MIT), where surprisingly Pr$_2$Ru$_2$O$_7$ is a Mott insulator and Ca$_2$Ru$_2$O$_7$ is a metal. This is unexpected because the valence-t$_{2\mathrm{g}}$ bands within the Ru-$4d^3$ manifold are half-filled for $A_2$Ru$_2$O$_7$, which naively should observe higher electronic correlation compared to the more than half-filled Ru-$4d^4$ bands in $R_2$Ru$_2$O$_7$ \cite{georges2013strong}. Moreover, Jiao \emph{et al.\ }\cite{jiao2018effect} demonstrated that Cd$_2$Ru$_2$O$_7$ is driven from metal to insulator by increasing pressure, in stark contrast to the related monoclinic compound Hg$_2$Ru$_2$O$_7$, which is a bad metal and becomes a good metal under pressure \cite{takeshita2007pressure}. Again, the behavior of Cd$_2$Ru$_2$O$_7$ is counter-intuitive, as without a structural transition the overlap integral between neighboring sites is expected to increase with pressure and, thus, electrons can more easily hop from one site to another, which increases their itinerancy. 

These experimental findings suggest a need to carefully treat strong electronic correlation effects in these systems. Thus, we will go beyond spin-density-functional theory \cite{kubler2017theory} (SDFT) and additionally employ the dynamical-mean-field theory \cite{georges1996dynamical} (DMFT). Yet, in order to truly remain a first-principles calculation, i.e., avoid free parameters, we obtain all the parameters used in the DMFT calculation by means of the constrained-random-phase approximation (cRPA) \cite{aryasetiawan2004frequency}. 

In the remainder of this paper, we focus on the magnetic structure and electronic properties of $R_2$Ru$_2$O$_7$ with $R^{3+}=$ Pr, Gd, Ho, and Er, as well as $A_2$Ru$_2$O$_7$ with nonmagnetic $A^{2+}=$ Ca, and Cd.  
First, we predict the magnetic ground state from first-principles devising the so-called cluster-multipole (CMP)+SDFT magnetic-structure-prediction scheme \cite{huebsch2021benchmark} and show that Ca and Cd compounds have robust all-in-all-out (AIAO) magnetic order within SDFT. Furthermore, for $R_2$Ru$_2$O$_7$, we find that different magnetic configurations are competing depending on the rare-earth element, which is consistent with experimental results. In particular, for Ho$_2$Ru$_2$O$_7$, we find a spin-ice-like state, while the remaining compounds prefer an antiferromagnetic (AFM) 32-pole structure.
Beyond that, cRPA calculations reveal that the relative energy of the O-$2p$ and O$'$-$2p$ bands \footnote{The O-$2p$ and O$'$-$2p$ orbitals are on the $48f$ and $8b$ site, respectively, as shown in \Cref{tab:1} and \Cref{fig:1}.} with respect to the Ru-\ttwog bands controls the screening of the on-site Coulomb repulsion $U$ and it differs significantly between Ru-4$d^3$ and Ru-4$d^4$ compounds. This explains the counter-intuitive reduction of the electronic correlation in the half-filled Ru-4$d^3$-\ttwog bands in $A_2$Ru$_2$O$_7$ compared to Ru-4$d^4$ bands in $R_2$Ru$_2$O$_7$. 
Lastly, we compute the electronic properties for Ca$_2$Ru$_2$O$_7$ and Pr$_2$Ru$_2$O$_7$ within DMFT using model parameters extracted from first-principles calculations by means of cRPA. The results show a bad metallic behavior for Ca$_2$Ru$_2$O$_7$ consistent with experimental results. Moreover, for Pr$_2$Ru$_2$O$_7$ a band gap opens, which is qualitatively and quantitatively in agreement with experimental results. 

\section{Magnetic structure} \label{sec:Magnetic structure}

\begin{table*}
    \caption{\label{tab:2} Basis configurations (BCs) of the cluster-multipole (CMP) basis for the cubic-pyrochlore structure $(R_{1-x}A_x)_2$Ru$_2$O$_6$O' with origin at $16c$ according to their multipole order and irreducible representation (IRREP). With Ru/$R_{1-x}A_x$ at positions of 
    Atom 1: $(0,0,0)$/ $(1/2,1/2,1/2)$, 
    Atom 2: $(1/4,1/4,0)$/ $(3/4,3/4,1/2)$, 
    Atom 3: $(1/4,0,1/4)$/ $(3/4,1/2,3/4)$, and 
    Atom 4: $(0,1/4,1/4)$/ $(1/2,3/4,3/4)$. The BCs are depicted in \Cref{fig:1}. Note that, the BCs of T$_{1\mathrm{g}}$ and T$_{2\mathrm{g}}$ are different domains of the same magnetic structure.
    }
    
    \begin{indented}
    \item[]\begin{tabular}{@{}lllllll@{}}
    \br
    Multipole & IRREP                    & BC          & Atom 1     & Atom 2    & Atom 3    & Atom 4 \\ \mr
    dipole    & $\Gamma_9$ (T$_{1\mathrm{g}}$)    & $\Psi_1$    & $(1,0,0)$  & $(1,0,0)$ & $(1,0,0)$ & $(1,0,0)$ \\ 
              &                          & $\Psi_2$    & $(0,1,0)$  & $(0,1,0)$ & $(0,1,0)$ & $(0,1,0)$ \\ 
              &                          & $\Psi_3$    & $(0,0,1)$  & $(0,0,1)$ & $(0,0,1)$ & $(0,0,1)$ \\ \mr
    octupole  & $\Gamma_3$ (A$_{2\mathrm{g}}$)    & $\sqrt{3}\Psi_4$    & $(-1,-1,-1)$  & $(1,1,-1)$ & $(1,-1,1)$ & $(-1,1,1)$ \\ \cline{2-7}
              & $\Gamma_9$ (T$_{1\mathrm{g}}$)  & $\sqrt{2}\Psi_5$    & $(0,1,1)$  & $(0,1,-1)$ & $(0,-1,1)$ & $(0,-1,-1)$ \\ 
                           &           & $\sqrt{2}\Psi_6$    & $(1,0,1)$  & $(1,0,-1)$ & $(-1,0,-1)$ & $(-1,0,1)$ \\ 
                           &           & $\sqrt{2}\Psi_7$    & $(1,1,0)$  & $(-1,-1,0)$ & $(1,-1,0)$ & $(-1,1,0)$ \\  \cline{2-7}
      &  $\Gamma_7$ (T$_{2\mathrm{g}}$) & $\sqrt{2}\Psi_8$    & $(0,-1,1)$  & $(0,-1,-1)$ & $(0,1,1)$ & $(0,1,-1)$ \\
                           &           & $\sqrt{2}\Psi_9$    & $(1,0,-1)$  & $(1,0,1)$ & $(-1,0,1)$ & $(-1,0,-1)$ \\
                           &           & $\sqrt{2}\Psi_{10}$ & $(-1,1,0)$  & $(1,-1,0)$ & $(-1,-1,0)$ & $(1,1,0)$ \\ \mr
    32 pole   &  $\Gamma_5$ (E$_{\mathrm{g}}$)  & $\sqrt{2}\Psi_{11}$ & $(-1,1,0)$  & $(1,-1,0)$ & $(1,1,0)$ & $(-1,-1,0)$ \\
                           &           & $\sqrt{6}\Psi_{12}$ & $(-1,-1,2)$  & $(1,1,2)$ & $(1,-1,-2)$ & $(-1,1,-2)$ \\
    \br
    \end{tabular}
    \end{indented}
\end{table*}

Let us start by developing some intuition about the competing magnetic interactions. As mentioned in the introduction, for $R_2$Ru$_2$O$_7$, the cubic-pyrochlore structure comprises two magnetically coupled frustrated sublattices at 16$c$ and 16$d$, that are dominated by $d$-orbital and $f$-orbital magnetism, respectively. This is shown in \Cref{fig:1} (a).

The $d$-orbital magnetism arises from the Ru-$4d^4$ bands in $R_2$Ru$_2$O$_7$, which host 4 electrons in the ionic limit. The exchange interaction between neighboring Ru moments is expected to be dominant compared to, e.g., dipole-dipole interactions. That is, because the on-site magnetic dipole moments on the Ru site are relatively small even if the high-spin state were realized, and the orbital-angular momentum $\mathbf{L}$ is likely quenched by the crystal-electric-field (CEF) splitting. Therefore, SOC is suppressed on the Ru site in all compounds investigated here. According to ligand-field theory \cite{sugano2012multiplets}, the overall cubic ($O_h$) symmetry splits the Ru-4$d$ bands into lower lying t$_{2\mathrm{g}}$ and upper e$_{\mathrm{g}}$ bands. The former is further split into a lower a$_{1\mathrm{g}}$ and upper e$'_\mathrm{g}$ bands due to the local trigonal ($D_{3d}$) symmetry. This yields a spin $S=1$ state in the Ru-$4d^4$ manifold for $R_2$Ru$_2$O$_7$, as shown in \Cref{fig:1} (b) bottom.

On the other hand, in the case of $f$-orbital magnetism, the electrons are more localized and, thus, better shielded from the surrounding CEF. Consequently, almost no quenching occurs and the heavy nuclei give rise to substantial SOC. One exception is given by the half-filled $f$-bands, where $\mathbf{L}=0$ for the ideal case of $R^{3+}$. In particular, in the compounds investigated here, SOC is expected to increase from $R^{3+}=$ Gd, Pr to Er and Ho. The ordering of the latter two cannot be predicted {\textit a priori} due to its dependence on multiple variables, whose values are expected to be of similar magnitude. For instance, SOC depends on the mass of the nuclei, the spin and angular contributions to the on-site magnetic moment. As shown in \Cref{fig:1} (c), the 4$f$ bands on the 16$d$-$R$ site observe a $D_{3d}$-site symmetry, which slightly lifts the local $2J+1$ degeneracy of the 4$f^n$ manifold. Here, $J$ is the total-angular-momentum-quantum number of $n$ electrons that occupy the 4$f$ orbitals on the $R$ site.

\red{The $R$ moment is expected to increases from Pr, Gd, Er to Ho when considering the $J$-$J$ coupling scheme, while the ionic radius ratio $r_R/r_{\mathrm{Ru}}$ decreases from Pr, Gd, Ho to Er~\footnote{\red{According to Figure 12 in Ref. \cite{subramanian1983oxide}, the values of the ionic radius ratio are approximately 1.81, 1.69, 1.63 and 1.61 for $R$=Pr, Gd, Ho and Er, respectively. In case of $A_2$Ru$_2$O$_7$, Figure 25 in Ref. \cite{subramanian1983oxide} yields 1.95 and 1.97 for $A$= Cd, and Ca.}}.}
Note that, the $d$-$f$ exchange interaction is expected to be large compared to $f$-$f$ exchange interaction due to the strong localization of the $f$ electrons. Intriguingly, the dipole-dipole interaction between $R$ moments may also play an important role between large on-site magnetic moments, as can emerge on the Ho and Er sites. This is seen in the related isostructural spin-ice compound Ho$_2$Tb$_2$O$_7$ \cite{harris1997geometrical,bramwell2001spin,bramwell2001science} and can lead to an effective ferromagnetic (FM) nearest-neighbor interaction, even if the exchange interaction is AFM.

In contrast, $A_2$Ru$_2$O$_7$ only features one magnetic site, i.e., the Ru site. 
The presence of $A^{2+}$ leads to a higher oxidation state Ru$^{5+}$ in the ionic limit. This in turn permits only $3$ electrons in the Ru-$4d^3$-a$_{1\mathrm{g}}$ and Ru-$4d^3$-e'$_\mathrm{g}$ bands, whose spins are expected to form a high-spin $S=3/2$ state according to Hund's rules, as illustrated in \Cref{fig:1} (b) top. This stands in contrast with $R_2$Ru$_2$O$_7$, where Ru$^{4+}$ yields an $S=1$ state in the Ru-$4d^4$ manifold. Consequently, even the undisturbed formation of magnetism on the Ru site is fundamentally different in $A_2$Ru$_2$O$_7$ compared to $R_2$Ru$_2$O$_7$. In other words, the magnetism on the Ru site in $A_2$Ru$_2$O$_7$ and $R_2$Ru$_2$O$_7$ must be discussed separately for two reasons: (i) the different state of oxidation, and (ii) the possible interaction with a second magnetic lattice formed by $R$ sites. 

\begin{figure}
  \includegraphics[width=\linewidth]{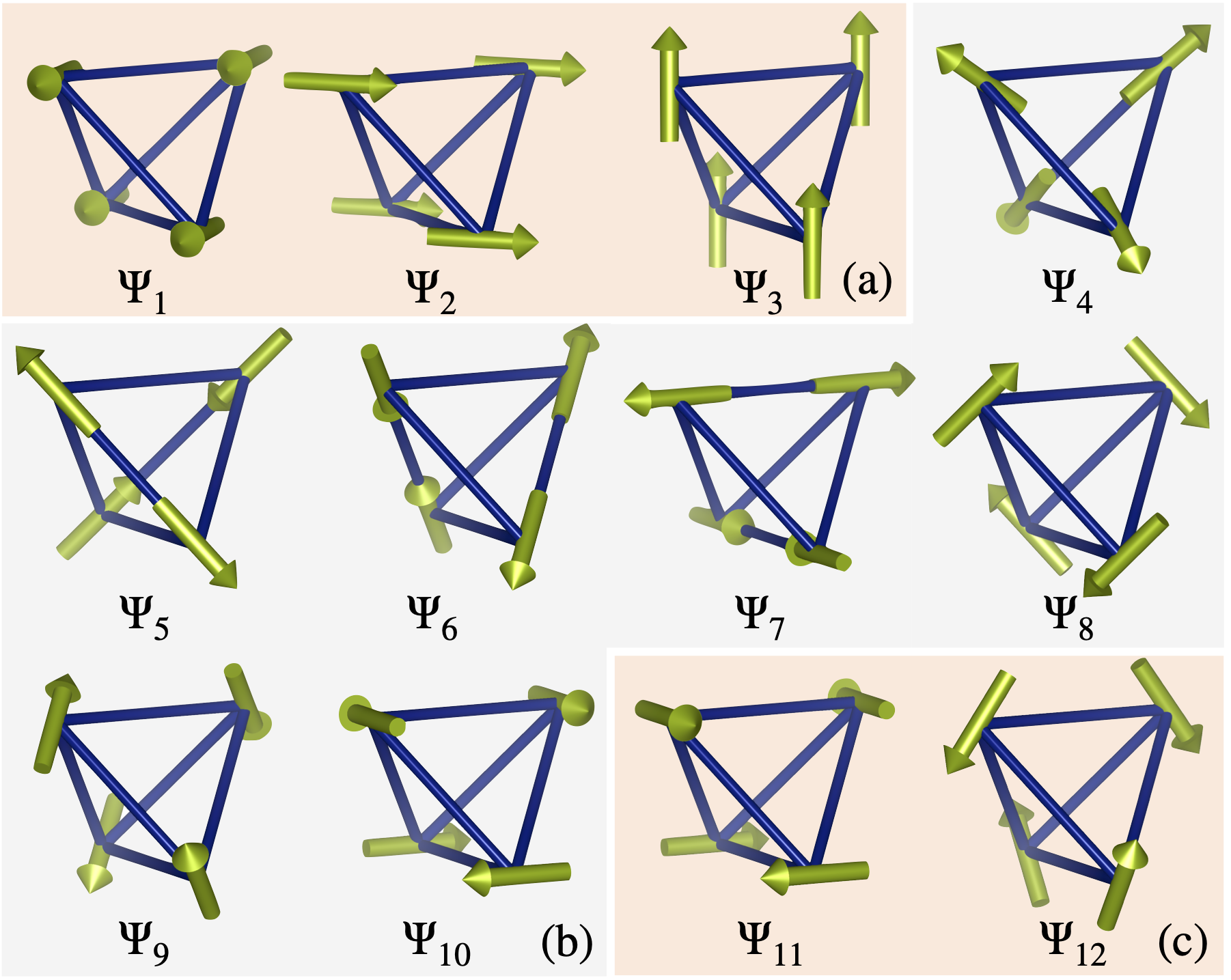}
  \caption{Cluster-multipole (CMP) basis. (a) Dipole basis configurations (BCs), (b) Octupole BCs, and (c) 32-pole BCs. As listed in \Cref{tab:2}, the BCs of the CMP basis are associated with an irreducible representation (IRREP), i.e., $\Psi_1$, $\Psi_2$, $\Psi_3$, as well as $\Psi_5$, $\Psi_6$, $\Psi_7$ are $\Gamma_9$(T$_{1\mathrm{g}}$), $\Psi_4$ is $\Gamma_3$(A$_{2\mathrm{g}}$), $\Psi_8$, $\Psi_9$, $\Psi_{10}$ are $\Gamma_7$(T$_{2\mathrm{g}}$), and  $\Psi_{11}$, $\Psi_{12}$ are $\Gamma_5$(E$_{\mathrm{g}}$). Note that, (i) $\Psi_4$ corresponds to the all-in-all-out (AIAO) structure, (ii) the spin-ice 2-in-2-out structure is a linear combination of dipole and octupole of IRREP $\Gamma_9$(T$_{1\mathrm{g}}$), e.g., $(\Psi_1+\Psi_5)/\sqrt{2}$.
  \label{fig:2}
  }
\end{figure}

Recently, magnetic multipoles are increasingly used to describe magnetic states in condensed matter on an atomic scale \cite{kusunose2008description,santini2009multipolar,hayami2018classification,hayami2018microscopic}, as well as on an inter-atomic scale \cite{dubovik1990toroid,hayami2016emergent,suzuki2017cluster,suzuki2018first,huyen2019topology,hayami2021essential}. This 
has led to the formulation of the so-called CMP theory \cite{suzuki2019multipole}, that proposes a basis to span the space of all possible magnetic configurations in a crystal in terms of magnetic multipoles. The magnetic configurations that form the CMP basis of cubic-pyrochlore ruthenates are explicitly given in \Cref{tab:2}. 

In \Cref{fig:2}, the 12 basis configurations (BCs) are depicted. These comprise dipoles, octupoles and 32 poles \footnote{We note that CMP theory can equally well define magnetic configurations corresponding to magnetic toroidal multipole moments. However, as the magnetic field is by construction an expansion of magnetic multipoles and we are not referring to effects where electrons couple to the gauge field, we choose to focus on magnetic multipoles here, in contrast to Ref.\ \cite{suzuki2019multipole}}. By construction, each BC additionally corresponds to a specific irreducible representation (IRREP) of the magnetic point group. 

It is well-known that most magnetic structures can be described with one IRREP, which can be understood in the context of Landau theory of second-order phase transitions \cite{landau2013course}. That is why Rietveld fits \cite{rietveld1969profile} to neutron diffraction patterns are usually performed with a basis similar to the CMP basis as a starting point \cite{hovestreydt1992karep,wills2000new,sikora2004mody,stokes2006isodisplace,aroyo2011crystallography,wills2015indexing}, which however lacks the characterization in terms of magnetic multipoles. The additional characterization of magnetic multipoles has two main advantages: First, it is rather intuitive that the complexity of the magnetic structure increases with the magnetic multipole order. Second, the shape of the linear response tensor can be directly inferred \cite{hayami2018classification}. For instance, the CMP theory has been instrumental in understanding the large anomalous Hall effect in the noncollinear AFM compound Mn$_3$Sn \cite{suzuki2017cluster,hayami2021essential}. 

Motivated by this success, some of the present authors developed a systematic scheme \cite{huebsch2021benchmark} to predict the magnetic ground state from first principles for a given crystal structure based on the CMP theory and SDFT. The predictive power of this scheme, that is termed CMP+SDFT \cite{huebsch2021benchmark}, has been demonstrated in a high-throughput calculation of more than 4400 SDFT calculations. 
The key issue, that was overcome, is that SDFT has many local minima in its total-energy landscape. An exhaustive list of candidate magnetic structures is instrumental to converge to all relevant (meta-)stable magnetic structures. To this end, we follow a statistically justified heuristic rule and take equally weighted linear combinations of all BCs with the same IRREP and multipole order. 
Note, that in the cubic-pyrochlore structure, the BCs of T$_{1\mathrm{g}}$ and T$_{2\mathrm{g}}$ are actually different domains of the same magnetic structure. In other words, they are related to each other by an alternative choice of the lattice vector and are degenerate by symmetry. That is, they will yield the same total energy in SDFT. 

After eradicating such redundant initial candidate magnetic configurations, we perform SDFT calculations using the Vienna \emph{ab-initio} simulation package (VASP) \cite{kresse1993ab,kresse1994ab,hobbs2001ab}. 
\red{We employ the pseudopotentials of the projector-augmented-waves (PAW) method version potpaw54 titled Ca\_sv, Cd, Pr, Gd, Ho, Er, O, and Ru\_pv in combination with the Perdew-Burke-Ernzerhof (PBE) exchange-correlation functional to converge the Kohn-Sham (KS) orbitals with the cutoff energy set to $520$\,eV in a self-consistency loop considering SOC \cite{steiner2016calculation} and noncollinear magnetism in the sense of J Kuebler's formulation of SDFT \cite{kubler2017theory}, where the KS Hamiltonian can be locally diagonalized. That is, the magnetic moments are not fixed during the self-consistent calculation.
We use a $\mathbf{k}$ mesh of $4\times4\times4$ devising the Monkhorst-Pack scheme and the structures that are listed in the inorganic crystal structure database \cite{ICSD_URL} under the ids 156409, 86773, 163397, 79332, 96730, and 97533. Note that we use the experimental crystal-structure information without performing additional ionic relaxation. It is provided along with the magnetic structure in the supplemental material. For more} details of the VASP calculations see Ref.\ \cite{huebsch2021benchmark}, where we used the same computational procedure.

\subsection{Magnetic structure of \texorpdfstring{$A_2$Ru$_2$O$_7$}{A2Ru2O7}} \label{subsec:Magnetic structure A}

After Cd$_2$Ru$_2$O$_7$ \cite{wang1998synthesis} in 1998, Ca$_2$Ru$_2$O$_7$ \cite{munenaka2006novel} could be synthesized in 2006. Finally, also Hg$_2$Ru$_2$O$_7$ \cite{yamamoto2007metal,klein2007hg} got synthesized in 2007, however it undergoes a structural phase transition \cite{van2012kagome} from cubic to monoclinic at the same time as the Ru moments order at 107K, which yields Kagome-like layers instead of corner-sharing tetrahedra. That is why, we focus on Ca$_2$Ru$_2$O$_7$ and Cd$_2$Ru$_2$O$_7$ in the CMP+SDFT calculations below..

While to our knowledge there is no experimental data available that directly probes the magnetic order in Ca$_2$Ru$_2$O$_7$ and Cd$_2$Ru$_2$O$_7$, some magnetic properties can be found in the literature.
The specific heat reveals distinct $\lambda$ anomalies at $T_N\approx85\,$K for Cd$_2$Ru$_2$O$_7$ \cite{jiao2018effect}, and at $T_N\approx107\,$K for Hg$_2$Ru$_2$O$_7$ \cite{klein2007hg}. These anomalies are associated with a second-order phase transition at which the Ru moments order. To our knowledge the specific heat for Ca$_2$Ru$_2$O$_7$ has not been reported. 
At zero-field no net magnetization is observed, so that the magnetic structures are all either AFM or glasslike. 

Further, the zero-field cooled (ZFC) magnetic susceptibility with $A=$ Cd \cite{miyazaki2010magnetic,jiao2018effect} and Hg \cite{klein2007hg} shows sharp cusps at $T_N$, and a speculative \emph{spin-glass} transition at $T_g\approx25\,$K for $A=$ Ca \cite{munenaka2006novel,taniguchi2009spin}. This further confirms that a magnetic transition occurs.  
For Cd$_2$Ru$_2$O$_7$ no high-temperature Curie--Weiss behavior is observed, but instead a broad maximum is stretched over a wide temperature range \cite{wang1998synthesis,jiao2018effect}. This is similar to CaCu$_3$Ru$_4$O$_{12}$, which can be discussed in terms of a high Kondo temperature which implies the presence of spin-fluctuations in a strongly-correlated-electron system \cite{krimmel2008non}. Additionally, the magnetic susceptibility of Cd$_2$Ru$_2$O$_7$ seems to reveal a second and third \emph{magnetic} transitions around $T_{m}\approx40\,$K and $T_{m2}\approx25\,$K \cite{jiao2018effect}, which are indicated by a sudden drop and a (reported, but not shown) hysteresis at $T_m$, as well as a very subtle kink around $T_{m2}$. Both compounds with $A=$ Cd, and Hg feature an increase of the ZFC magnetic susceptibility with decreasing temperature, and the nearly identical behavior in the field-cooled (FC) magnetic susceptibility.

For Ca$_2$Ru$_2$O$_7$ the size of the effective Ru moment could be inferred from a Curie--Weiss fit to the high-temperature magnetic susceptibility, which yields a surprisingly small value of $0.25-0.36\mub/$Ru \cite{munenaka2006novel,taniguchi2009spin}. As a comparison, the fully spin-polarized value would be $3.87\mub/$Ru using $S=3/2$ as the spin-quantum number. 
And even the low-spin state yields $1.73\mub/$Ru.
Moreover, the magnetic susceptibility of Ca$_2$Ru$_2$O$_7$ shows a pronounced split between ZFC and FC curve \cite{munenaka2006novel}, with no upturn in the ZFC case and an almost constant FC measurement. 
In fact, Taniguchi \emph{et al.\ }\cite{taniguchi2009spin} have first ascribed a thermodynamic spin-glass transition to Ca$_2$Ru$_2$O$_7$ at $T_g$ by measuring higher-order magnetic susceptibilities.

A spin-glass state is characterized by random, localized magnetic moments with a slow response to external magnetic fields. The muon-spin-rotation ($\mu$SR) measurements by Miyazaki \emph{et al.\ }\cite{miyazaki2010magnetic} give further insights to the spin-dynamic properties of $A_2$Ru$_2$O$_7$ with $A=$ Cd, Ca, and Hg. They report that around the MIT, a nearly commensurate magnetic order develops in $A=$ Hg, and Cd. But in Cd$_2$Ru$_2$O$_7$ below $T_m$, the presence of a Gaussian distribution at the muon site is interpreted as some randomness of the Ru moments on a regular lattice. For Ca$_2$Ru$_2$O$_7$, the same kind of randomness is most pronounced and dubbed \emph{frozen spin liquid}.  
According to the $\mu$SR results, the size of the effective Ru moments is $0.36(7)\mub/$Ru for Cd$_2$Ru$_2$O$_7$ for $T_m<T<T_N$, $0.60\mub/$Ru (for $T < T_g$, but $0.35\mub/$Ru for $T > T_g$) for Ca$_2$Ru$_2$O$_7$, and $0.5(1)\mub$/Ru for Hg$_2$Ru$_2$O$_7$.

\begin{table*}
    \caption{\label{tab:3}CMP+SDFT results for $A_2$Ru$_2$O$_7$. The table lists the magnetic space group (MSPG), Laue group, total energy with respect to the AIAO structure ($\Psi_4$, highlighted in orange), and predicted size of the magnetic moment along the quantization axis of CMP+SDFT (meta-)stable magnetic structures below $10\,$meV and with zero net magnetization.}
    \begin{indented}
    \item[] \begin{tabular}{lllll}
    \br
    &        MSPG     & Laue group    & meV  & $\mub/$Ru  \\ \mr
\cellcolor{grey}Ca$_2$Ru$_2$O$_7$ &      &    &   &   \\
\rowcolor{orange}$\Psi_4$                                  & Fd$\bar{3}$m'(227.131)  &  m-3m'      & 0.00  & 1.17  \\
$\Psi_4 \Psi_5 \Psi_7 \Psi_8 \Psi_{10}$   & C2'/m'(12.62)           &  2'/m'      & 3.96  & 1.10  \\
$\Psi_8$                                  & I4$_1$'/amd'(141.555)   &  4'/mmm'    & 5.59  & 1.12  \\
$\Psi_5 \Psi_8 $                          & Fd'd'd(70.530)          &  m'm'm      & 5.99  & 1.09  \\
$\Psi_4 \Psi_{12} $                       & I4$_1$'/am'd(141.554)   &  4'/mm'm    & 8.51  & 1.08  \\
$\Psi_5 $                                 & C2'/c'(15.89)           &  2'/m'      & 9.46  & 1.08  \\
\mr
\cellcolor{grey}Cd$_2$Ru$_2$O$_7$ &      &    &   &   \\
\rowcolor{orange} $\Psi_4$                                  & Fd$\bar{3}$m'(227.131)  & m-3m'      & 0.00  & 1.07  \\
$\Psi_8$                                  & I4$_1$'/amd'(141.555)   &  4'/mmm'    & 2.47  & 0.93  \\
$\Psi_5 \Psi_7 \Psi_8 \Psi_{10}$          & Imm'a'(74.559)          &  m'm'm      & 3.17  & 0.87  \\
$\Psi_5 \Psi_6 \Psi_7$                    & R$\bar{3}$m'(166.101)   &  -3m'1      & 5.36  & 0.77  \\
$\Psi_5$                                  & I4$_1$/am'd'(141.557)   &  4/mm'm'    & 6.96  & 0.77  \\
\br
    \end{tabular}
    \end{indented}
\end{table*}

Our CMP+SDFT calculations predict that $\Psi_4$ is the most stable magnetic structure for both, Ca$_2$Ru$_2$O$_7$ and Cd$_2$Ru$_2$O$_7$, as seen in \Cref{fig:2} (a). Note that, all Ru moments in one tetrahedron point outward, which implies that the corner-sharing tetrahedra comprise Ru moments pointing inward. Thus, the magnetic ground state is dubbed AIAO structure. It is separated by only a few meV in total energy from other metastable magnetic configurations obtained by CMP+SDFT. The occurrence of almost degenerate local minima in SDFT is in good agreement with the experimental observation of multiple magnetic transitions. 

A list of the magnetic space group (MSPG), Laue group, total energy with respect to the AIAO structure and the predicted size of the Ru moment for all local minima is presented in \Cref{tab:3}. The first column states the dominant BCs, although smaller contributions from additional BCs might further break the symmetry. For the AIAO structure, the MSPG is Fd$\bar{3}$m' and the predicted size of the Ru moments within SDFT is $1.17\mub/$Ru for Ca$_2$Ru$_2$O$_7$, and $1.07\mub/$Ru for Cd$_2$Ru$_2$O$_7$. Note that here we refer to the Ru moment along the quantization axis considering both contributions, the spin and angular-momentum contribution, though the latter is negligible.
The predicted value is much small than the high-spin state in the ionic limit, i.e., $3\mub$, which implies that the system disobeys Hund's rules as a result of CEF splitting between e$'_\mathrm{g}$ and $a_{1\mathrm{g}}$ bands due to $D_{3d}$-site symmetry. Nevertheless, the Ru moments are much larger than the experimentally observed value. This might be due to electronic correlation effects which are not captured in the SDFT-ground-state calculation and will be discussed in \Cref{sec:Electronic properties}.

The corresponding magnetic crystal information files (MCIFs) for all magnetic structures given in \Cref{tab:3} are provided in the supplemental material. From \Cref{tab:3}, we see that the SDFT-total-energy landscapes of Ca$_2$Ru$_2$O$_7$ and Cd$_2$Ru$_2$O$_7$ are similar in the sense that (i) the most stable configuration is the AIAO structure given by $\Psi_4$, (ii) $\Psi_8$ with MSPG I4$_1$'/amd'(141.555) is a local minimum, (iii) magnetic structures dominated by $\Psi_5$ are local minima, (iv) linear combinations of octupole BCs form other similar local minima. As the AIAO structure has the lowest total energy among all CMP+SDFT calculations and is consistent with the net zero magnetization observed experimentally, it is used in the cRPA and DMFT calculations in \Cref{sec:Electronic properties}.

\begin{figure}
  \includegraphics[width=\columnwidth]{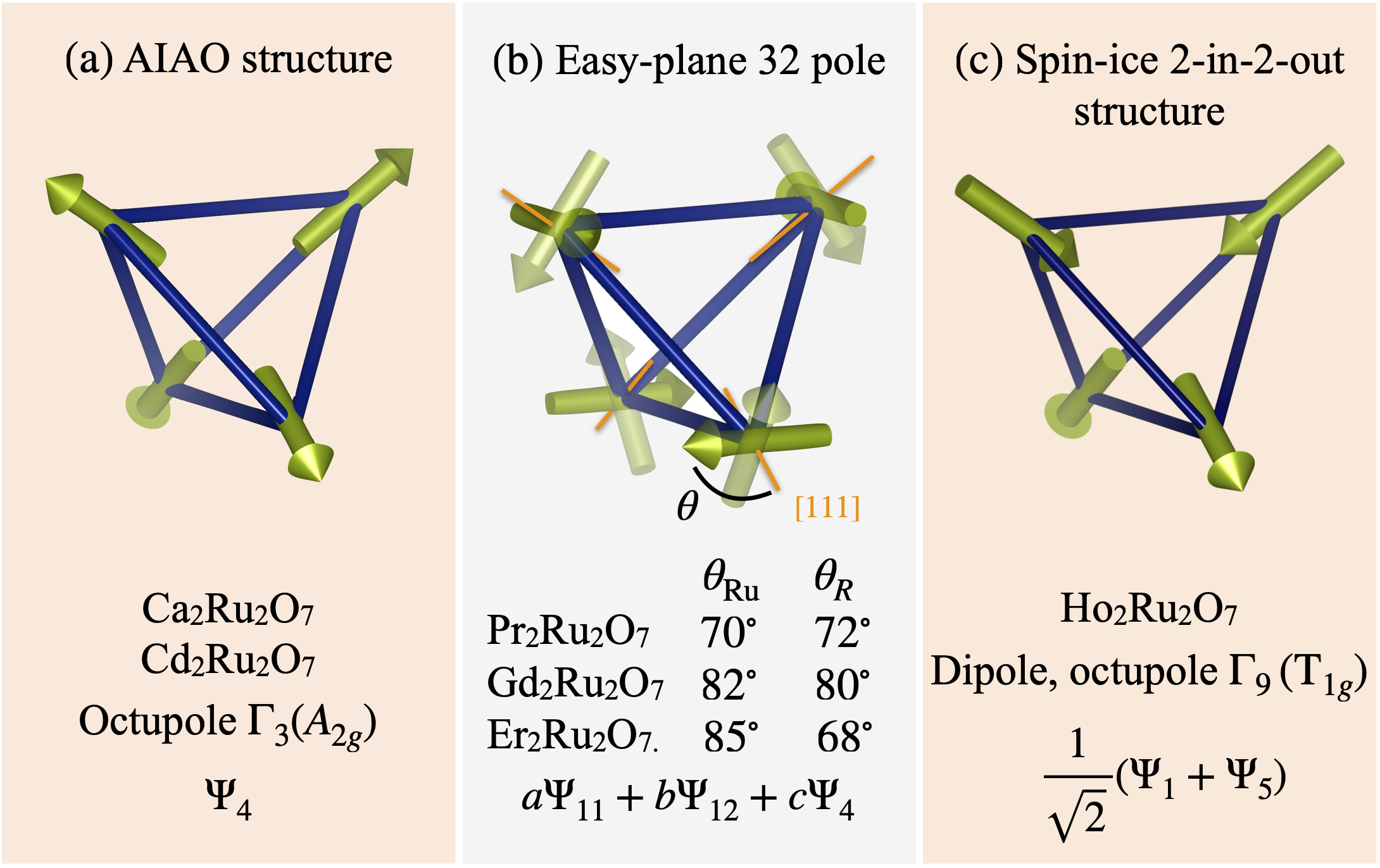}
  \caption{(a) Antiferromagnetic all-in-all-out (AIAO) structure, which corresponds to an octupole with irreducible representation $\Gamma_3(A_{2g})$ and $\Psi_4$ in \Cref{tab:2}. It is the CMP+SDFT ground state on the Ru sublattice in Ca$_2$Ru$_2$O$_7$ and Cd$_2$Ru$_2$O$_7$. (b) Easy-plane 32 pole with $\theta=90^{\circ}$. A finite contribution of $\Psi_{4}$ varies $\theta$. (c) Spin-ice 2-in-2-out structure.
  \label{fig:3}
  }
\end{figure}

\subsection{Magnetic structure of \texorpdfstring{$R_2$Ru$_2$O$_7$}{R2Ru2O7}} \label{subsec:Magnetic structure R}

Let us now consider cubic-pyrochlore rare-earth ruthenates, $R_2$Ru$_2$O$_7$. The Ru site hosts a spin $S=1$ state in the a$_\mathrm{1g}$ and e$'_\mathrm{g}$ bands, and is, consequently, fundamentally different from the Ru site in $A_2$Ru$_2$O$_7$ discussed in the previous section. For instance, owing to the integer value of the spin, the system can in principle be in a quantum-paramagnetic phase with $\mathbf{S} \cdot \hat{z} = S_z = 0$ \cite{li2018competing}, where $\hat{z}$ is the local $[111]$ direction along the $D_{3d}$ axis. The magnetic ground state of the Ru moments depends on the O--Ru bond length and, thus, on the ionic radius of the rare-earth element $R=$Pr, Gd, Ho, and Er. Therefore, it depends on $R$ even before considering the $d$-$f$-exchange interaction. For the size of the on-site magnetic dipole moment on the Ru site, the ionic limit is given by $2.83\mub$/Ru using $S=1$. Again, the on-site moment for the $R$ site varies depending on the element.

\red{The selected rare-earth elements cover a wide range of different prevalent mechanisms. In particular, for Pr the spin and orbital contributions have a comparable size and combine antiferromagnetically. Thus, the Pr moments are expected to be relatively small and well shielded. The ordering should emerge predominantly under the influence of spin-orbit coupling. On the other hand, for Gd the $R$-$4f$ bands are half-filled, which results in a vanishing orbital angular momentum and, thus, negligible SOC. Here, $d$-$f$-exchange interaction and $f$-$f$-exchange interaction could be most important. In Er$_2$Ru$_2$O$_7$ and Ho$_2$Ru$_2$O$_7$ the spin and orbital contributions add up ferromagnetically which amounts to a large $R$ moment in both cases. In contrast to the case of Pr, these are promising candidate compounds for a dominant dipole-dipole interaction amongst neighboring $R$ sites.}

In a broad survey, Ito \emph{et al.\ }\cite{ito2001nature} reported the experimental temperature dependence of the specific heat of $R_2$Ru$_2$O$_7$. In the investigated temperature range, second-order phase transitions associated with the ordering of Ru moments are observed at $T_N\approx160$K for Pr$_2$Ru$_2$O$_7$ \cite{taira1999magnetic,ito2001nature,tachibana2007heat,van2017induced}, $114$K for Gd$_2$Ru$_2$O$_7$ \cite{ito2001nature,gurgul2007bulk}, $95$K for Ho$_2$Ru$_2$O$_7$ \cite{ito2001nature,wiebe2004magnetic}, and $90$K for Er$_2$Ru$_2$O$_7$ \cite{ito2001nature,taira2003magnetic}. In contrast to $A_2$Ru$_2$O$_7$, for $R_2$Ru$_2$O$_7$, possible magnetic structures have been recently reported based on neutron scattering using powder diffraction.

The magnetic structures predicted by CMP+SDFT are summarized in \Cref{tab:4}. The first column lists the most dominant BCs, where $\Psi'$ corresponds to the sublattice of $R$ sites and $\Psi$ to Ru sites. Note that, the CMP basis for the sublattice of both, $R$ sites and Ru sites, is identical and presented in \Cref{fig:2} and \Cref{tab:2}. Furthermore, \Cref{tab:4} lists the MSPG, Laue group, the total energy with respect to the minimum total energy identified within CMP+SDFT ($E_{\mathrm{tot}}$ in meV), size of the on-site magnetic dipole moment on the $R$ site and Ru site considering both contributions, the spin and angular-momentum contribution, along the quantization axis,
as well as the net magnetization $\mathbf{M}$ per unit formula. 
In the following, we will discuss the agreement of our numeric results with known experimental observations individually for each compound.

\begin{table*}
    \caption{CMP+SDFT results. The table lists the dominant basis configurations $\Psi_i$ ($\Psi'_i$) for the Ru ($R$) sublattice, magnetic space group (MSPG), Laue group, total energy $E_{\mathrm{tot}}$ with respect to the most stable magnetic structure among all CMP+SDFT calculations, predicted size of the magnetic moment of the Ru and $R$ site along the quantization axis, and the net magnetization per unit formula $\mathbf{M}$ for $R_2$Ru$_2$O$_7$. For Gd$_2$Ru$_2$O$_7$ and Ho$_2$Ru$_2$O$_7$ only results with $E_{\mathrm{tot}}\leq10\,$meV are reported. In case of Gd$_2$Ru$_2$O$_7$ only (meta-)stable states with zero net magnetization are listed. For Pr$_2$Ru$_2$O$_7$ and Er$_2$Ru$_2$O$_7$ results with $E_{\mathrm{tot}}\leq10\,$meV and additionally local minima with $E_{\mathrm{tot}}>10\,$meV and zero net magnetization are reported. In orange we highlight the magnetic order we expect is most stable under the constraints known from experimental data as discussed in the main text. 
    \label{tab:4}
    }
    \lineup
    \begin{indented}
    \item[]\begin{tabular}{lllllll}
    \br
    &        MSPG     & Laue group    & meV  & $\mub/R$ & $\mub/$Ru & $\mathbf{M}$($\mub/$u.f.) \\ \mr 
\cellcolor{grey}Pr$_2$Ru$_2$O$_7$ &      &    &   & & & \\
$\Psi'_1 \Psi'_3 \Psi'_5 \Psi'_7 \Psi'_8 \Psi'_{10} \Psi_1 \Psi_3 \Psi_5 \Psi_7 \Psi_8 \Psi_{10}  $        & Imm'a'(74.559)        &  m'm'm         & \0\00.00  & 0.80 & 0.94  & \06.00 \\
$\Psi'_1 \Psi'_2 \Psi_5 \Psi_8 \Psi_{11} $                                                                 & P$\bar{1}$(2.4)       &  $\bar{1}$     & \0\03.79  & 0.76 & 0.95  & \06.22 \\
$\Psi'_1 \Psi'_2 \Psi'_3 \Psi_5 \Psi_6 \Psi_7 $                                                            & R$\bar{3}$m'(166.101) &  $\bar{3}$m'1  & \0\05.18  & 0.74 & 0.91  & \06.01 \\
\rowcolor{orange}$\Psi'_4 \Psi'_{12} \Psi_4 \Psi_{12} $                                                    & I4$_1$'/am'd(141.554) &  4'/mm'm       & 142.19  & 0.77 & 1.03  & \00.00 \\
\rowcolor{orange}$\Psi'_{11} \Psi_{11} $                                                                   & I4$_1$/amd(141.551)   &  4/mmm         & 151.00  & 1.01 & 0.90  & \00.00 \\
$\Psi'_9 \Psi_9 $                                                                                          & I4$_1$'/amd'(141.555) &  4'/mmm'       & 151.31  & 1.19 & 0.88  & \00.00 \\
\mr
\cellcolor{grey}Gd$_2$Ru$_2$O$_7$ &      &    &  &  & &  \\
\rowcolor{orange} $\Psi'_4 \Psi'_{11} \Psi'_{12} \Psi_4 \Psi_{11} \Psi_{12} $                    & Fddd(70.527)           &  mmm          & \0\00.00  & 6.86 & 1.28  & \00.00 \\
$\Psi'_{11} \Psi'_{12} \Psi_{11} \Psi_{12} $                                   & Fddd(70.527)           &  mmm          & \0\07.69  & 6.85 & 1.28  & \00.00 \\
$\Psi'_{11} \Psi_{11} $                                                        & I4$_1$/amd(141.551)    &  4/mmm        & \0\07.80  & 6.84  & 1.28  & \00.00 \\
\mr
\cellcolor{grey}Ho$_2$Ru$_2$O$_7$ &      &    &   & &  \\
\rowcolor{orange} $\Psi'_1 \Psi'_5 (\Psi_1) \Psi_5 $                                             & I4$_1$/am'd'(141.557)  &  4/mm'm'       & \0\00.00  & 9.20 & 1.29  & 10.40 \\
$\Psi'_4 \Psi_4 $                                                              & Fd$\bar{3}$m'(227.131) &  m$\bar{3}$m'  & \0\02.17  & 9.26 & 1.19  & \00.00 \\
\mr
\cellcolor{grey}Er$_2$Ru$_2$O$_7$ &      &    &  &  & &  \\
$\Psi'_3 \Psi'_7 \Psi_3 \Psi_7 $                                               &  I4$_1$/am'd'(141.557) &  4/mm'm'       & \0\00.00  & 7.07 & 1.26  & \08.54 \\
$\Psi'_1 \Psi'_5 (\Psi'_{11} \Psi'_{12}) (\Psi_1 \Psi_5) \Psi_{11} \Psi_{12} $ & I4$_1$/a(88.81)        &  4/m           & \0\00.00  & 7.07 & 1.29  & \08.25 \\
$\Psi'_1 \Psi'_5 (\Psi'_{12}) \Psi_5 \Psi_{11} \Psi_{12}$                      & I4$_1$/a(88.81)        &  4/m           & \0\09.14  & 7.08 & 1.19  & \08.81 \\
\rowcolor{orange} $\Psi'_{11} \Psi_{11} $                                                        & I4$_1$/amd(141.551)    &  4/mmm         & \038.84 & 7.57 & 1.36  & \00.00 \\
$\Psi'_{11} \Psi_{11} $                                                        & I4$_1$/amd(141.551)    &  4/mmm         & \051.72  & 7.54 & 1.30  & \00.00 \\
$\Psi'_{11} \Psi'_{12} \Psi_8 $                                                & Imma(74.554)           &  mmm           & \056.76  & 7.55 & 1.35  & \00.00 \\
$\Psi'_4 \Psi'_{12} (\Psi_4) \Psi_{12} $                                       & I4$_1$'/am'd(141.554)  &  4'/mm'm       & \074.08  & 7.08 & 1.30  & \00.00 \\
$\Psi'_4 \Psi_4 $                                                              & Fd$\bar{3}$m'(227.131) &  m$\bar{3}$m'  & 102.70  & 6.99 & 1.14  & \00.00 \\
\br
    \end{tabular}
    \end{indented}
\end{table*}

\subsubsection{\texorpdfstring{Pr$_2$Ru$_2$O$_7$}{Pr2Ru2O7}} \white{Without this visibility is very low.}
\newline
In particular for Pr$_2$Ru$_2$O$_7$, Van Dunijn \emph{et al.\ }\cite{van2017induced} determined the magnetic order of the Ru sublattice to be a linear combination of $\Psi_{11}$ and $\Psi_{12}$ of the CMP basis shown in \Cref{fig:3} (b) with $c=0$, i.e., $\theta_{\mathrm{Ru}}=90{^\circ}$. Hence, that magnetic structure is a 32 pole in terms of the CMP theory. Regrettably, it is not possible to identify which linear combination of $\Psi_{11}$ and $\Psi_{12}$ is realized based on the powder-diffraction measurement due to the intrinsic ambiguity of the direction of the Ru moments in the $(111)$ plane perpendicular to the local $\hat{z}$ axis, i.e., the $[111]$ direction, when using a powdered sample. In other words, $a$ and $b$ in \Cref{fig:3} (b) are unknown. Nevertheless, it is clear that the Ru moments possess AFM order in the local $\hat{x}\hat{y}$ plane, which is distinct from being a quantum paramagnet. The same magnetic structure has been experimentally inferred for the Ru sites of Y$_2$Ru$_2$O$_7$ \cite{ito2001nature}, and for the Er sites of Er$_2$Ti$_2$O$_7$ \cite{poole2007magnetic}. The latter is said to feature an accidental continuous rotational symmetry in the $\hat{x}\hat{y}$ plane, which is lifted with an order-by-disorder mechanism \cite{savary2012order}. Moreover, Pr is reported not to order down to $100\,$mK \cite{van2017induced} and no more than $0.(3)\mub$/Pr may be present without worsening the fit to the experimental data. 

The magnetic structure with the lowest total energy in SDFT features BCs with $\Gamma_9$(T$_{1g}$) and $\Gamma_9$(T$_{2g}$). It has a finite net magnetization in contrast to the experimental observation. The size of the magnetic moments on both sites, Ru and Pr, is comparable to each other in SDFT. Experimentally though, the Ru moments are reported to be $1.48(4)\mub$/Ru \cite{van2017induced}, 
which is a reduction compared to the ionic limit, i.e., $2\mub$.  We note that the size of Ru moments predicted by SDFT is underestimated compared to the experiment, $0.94\mub$/Ru $< 1.48(4)\mub$/Ru, but almost within the accuracy of $\pm0.5\mub$ expected for SDFT \cite{huebsch2021benchmark}. On the other hand, the size of Pr moments seems to be overestimated by SDFT, $0.80\mub$/Pr $> 0.(3)\mub$/Pr. 

In the SDFT calculations, it is mostly the orbital contribution of the Pr-$f$ electrons that introduces a FM tendency towards $\Psi'_1$ and $\Psi'_3$ with $\Gamma_9$(T$_{1\mathrm{g}}$). In contrast, the Ru order is rather biased towards $\Psi_5$ and $\Psi_7$ with $\Gamma_9$(T$_{1\mathrm{g}}$), as well as $\Psi_8$ and $\Psi_{10}$ with $\Gamma_9$(T$_{2\mathrm{g}}$). Note that, both pairs of BCs are octupoles, but with different IRREPs, c.f., \Cref{tab:2}. In \Cref{tab:4}, we see that Pr sites tend to order ferromagnetically with BCs of IRREP $\Gamma_9$(T$_{1\mathrm{g}}$) also in other low-lying local minima with $E_{\mathrm{tot}} < 10\,$meV, which seems to induce octupole order of the same IRREP on the Ru sites. As SDFT overestimates the size of Pr moments and its order influences the order at the Ru sites, we infer that the tendency toward FM order is also overestimated.
That is why the low-lying local minima are not reproducing the experimentally observed AFM order.

Nevertheless, the CMP+SDFT scheme is able to find local minima with zero net magnetization at higher total energy. And indeed, the magnetic structures of these AFM local minima are dominated by BCs corresponding to a 32 pole. In particular, the 32 pole highlighted in \Cref{tab:4} is more stable than other AFM local minima with octupole order. A contribution of $\Psi'_{4}$ and $\Psi_{4}$ introduces an angle $\theta_{\mathrm{Ru}}\approx70^{\circ}$ and $\theta_{\mathrm{Pr}}\approx72^{\circ}$ as discussed in more detail for Gd$_2$Ru$_2$O$_7$ in the next subsection. We conclude that, we could find an easy-plane 32-pole AFM structure to be among the most stable AFM configurations, which is in good agreement with the experimental result. However, the size of the Pr moments is overestimated within SDFT, which introduces an FM tendency that is in conflict with the experimental observation.

\subsubsection{\texorpdfstring{Gd$_2$Ru$_2$O$_7$}{Gd2Ru2O7}} \white{Without this visibility is very low.}
\newline
For Gd$_2$Ru$_2$O$_7$, Mössbauer spectroscopy shows that Ru moments order at $T_N\approx114$K \cite{gurgul2007bulk}. In the magnetic susceptibilities \cite{gurgul2007bulk,yao2011hydrothermal}, there is a difference between FC and ZFC with a small history-dependent component, which might be explained by a tiny FM contribution due to a canting of the AFM structure. According to Gurgul \emph{et al.\ }\cite{gurgul2007bulk}, the Ru moments order almost as a 32 pole, but the angle $\theta_{\mathrm{Ru}}$ enclosed by the Ru moments and the $\hat{z}$ axis is not perfectly $90^{\circ}$ as illustrated in \Cref{fig:3} (b). Instead, $\theta_{\mathrm{Ru}}$ is reported to be $72^{\circ}$ \cite{gurgul2007bulk}.
This can be accounted for by a linear combination of the BCs of the 32 pole and the octupole $\Psi_4$. The resulting magnetic structure is extraordinary in the sense that, it features a combination of multiple IRREPs and this is rather unusual in the context of commonly experimentally determined magnetic structures \cite{huebsch2021benchmark}. Here, the order is a combination of $\Psi_{4}$ with $\Gamma_3$(A$_{2g}$), and $\Psi_{11}$ and $\Psi_{12}$ with $\Gamma_5$(E$_g$), c.f., \Cref{tab:2}, \Cref{fig:2} (b), (c) and \Cref{fig:3} (b). And indeed, a linear combination of the BCs of the 32 poles, $\Psi_{11}$ and $\Psi_{12}$, and the octupole $\Psi_4$ is most stable on both, the Gd and Ru sublattices, also within the CMP+SDFT scheme as highlighted in \Cref{tab:4}. The angle $\theta_{\mathrm{Ru}}\approx82^{\circ}$, which is closer to $90^{\circ}$ than the experimental result ($72^{\circ}$), but reproduces the overall tendency.

Let us note that, the IRREP of the dipole is $\Gamma_9$(T$_{1g}$). Therefore, neither $\Psi_{11}$ and $\Psi_{12}$ nor $\Psi_4$ couples directly to either of the dipole BCs. This stands in contrast to the octupole-magnetic-ground-state structure in Mn$_3$Sn, where the octupole and dipole have the same IRREP and directly couple, which explains the large anomalous response observed in this compound. That means, the small history dependence of the magnetic susceptibility remains puzzling for Gd$_2$Ru$_2$O$_7$. However, besides the AFM local minima listed in \Cref{tab:2}, we identified the staggering number of 16 local minima for Gd$_2$Ru$_2$O$_7$ with $E_{\mathrm{tot}} < 10$meV. These have a finite net magnetization and are not listed explicitly in \Cref{tab:2}. Let us mention that many of these metastable magnetic structures feature BCs with IRREP $\Gamma_9$(T$_{1g}$), which easily couple to an applied magnetic field. 

At $40$K, also the Gd orders as a 32 pole \cite{gurgul2007bulk}, but again no conclusion can be drawn concerning the directions of the Gd moments in the $\hat{x}\hat{y}$ plane based on the experimental data. To our knowledge, it has not been confirmed whether a second-order phase transition can be discerned from the specific heat \cite{taira2002magnetic} around $40$K. 
We find that also the Gd sublattices feature a finite angle $\theta_{\mathrm{Gd}}\approx 80^{\circ}$, which was not seen in the experiment. We emphasize that (i) all three AFM local minima below $10$meV feature a 32-pole order, and (ii) the exact linear combination of $\Psi_{11}$ and $\Psi_{12}$ is quasi degenerate as seen by comparing the two metastable magnetic structures around $E_{\mathrm{tot}} = 7.7$meV.

The size of the effective Ru moments is reported to be $2.58 \mub$/Ru \cite{taira2002magnetic}, or $2.12 \mub$/Ru \cite{yao2011hydrothermal}, where the former is close to the fully saturated effective Ru moment $2.83\mub$. In contrast, we find the Ru moment along the quantization axis is $1.28\mub$/Ru within SDFT, which is a reduction by $36\%$ compared to the saturated value of $2\mub$/Ru.
For the Gd site, we could not find any report on the size of the Gd moments. Nevertheless, the predicted size of the Gd moments in SDFT seems reasonable when compared to the experimental value for other materials, such as GdB$_4$ with $7.14\mub$/Gd \cite{Blanco2006} and GdVO$_4$ \cite{Palacios2018} with $7.0\mub$/Gd.

\subsubsection{\texorpdfstring{Ho$_2$Ru$_2$O$_7$}{Ho2Ru2O7}} \white{Without this visibility is very low.}
\newline
Ho$_2$Ru$_2$O$_7$ has the most intriguing interplay between $d$ and $f$-electron magnetism and features spin-ice-like states. 
As shown in \Cref{fig:3} (c), a perfect spin-ice structure has two magnetic moments pointing inward along the local $\hat{z}$ axis and two pointing outward in each tetrahedron. It can be obtained by a linear combination of $\Psi_1$ and $\Psi_5$, which both have the same IRREP as seen in \Cref{tab:2}. The Ru moments in Ho$_2$Ru$_2$O$_7$ order close to that 2-in-2-out spin-ice structure around $95$K \cite{wiebe2004magnetic,gardner2005spin}. However compared to other known spin ice, e.g., Ho$_2$Ti$_2$O$_7$ \cite{bramwell2001science}, it behaves as if an external magnetic field of approximately $1\,$T is present \cite{gardner2005spin}. This has been experimentally confirmed by measurements of the magnetic entropy. Interestingly, the low-temperature specific heat \cite{gardner2005spin} has both (i) a broad feature around $3\,$K associated with the freezing of magnetic moments, and (ii) a sharp $\lambda$ anomaly below $2$K indicating a second-order phase transition.

At $T_{N}\approx1.4$K \cite{wiebe2004magnetic}, the Ho moments order in a spin-ice–like state, but with an additional long-ranged FM order between neighboring tetraheadra within the Ho sublattice due to a small canting of the Ho moments. According to Wiebe \emph{et al.\ }\cite{wiebe2004magnetic}, the Ru moments then abandon the spin-ice-like state and form a nearly collinear FM in the low-temperature regime of $100\,$mK. This fully compensates the FM order of the Ho sublattice and yields zero net magnetization, so that the magnetic structure of Ho$_2$Ru$_2$O$_7$ is overall AFM.

In CMP+SDFT, the ground state is a spin-ice-like order on Ho sites, where almost equal contributions from $\Psi'_1$ and $\Psi'_5$ emerge. On the Ru sites, we find rather $\Psi_5$ with little contribution from $\Psi_1$. This is the opposite of the experimentally reported behavior, where Ru sites order ferromagnetically to compensate the Ho moments. Therefore, we observe a finite net magnetization in our calculations. 
\red{Moreover, CMP+SDFT predicts the AIAO structure to be a second almost degenerate metastable magnetic configuration. This hints towards a fickle balance between ferromagnetic and antiferromagnetic interactions on the geometrically frustrated lattice and it might be easily possible to tune the magnetic ground state to obtain the AIAO structure by changing external conditions, e.g., by applying pressure. }

The size of the Ru moment is reported as $1.2(2) \mub$/Ru \cite{wiebe2004magnetic}, and the effective Ru moment is reported by different authors as $3.59 \mub$/Ru \cite{yao2011hydrothermal}, and $4.32 \mub$/Ru \cite{taira2002magnetic}. Moreover, the size of the effective Ho moment is reported as $9.60(1) \mub$/Ho \cite{bansal2003magnetic} from a Curie-Weiss fit, and the size of the Ho moment in neutron diffraction is $9.29(3) \mub$/Ho \cite{wiebe2004magnetic}.
The size of Ru and Ho moments predicted within SDFT agree well with the values reported by Wiebe \emph{et al.\ }\cite{wiebe2004magnetic}.

\subsubsection{\texorpdfstring{Er$_2$Ru$_2$O$_7$}{Er2Ru2O7}} \white{Without this visibility is very low.}
\newline
For Er$_2$Ru$_2$O$_7$, Taira \emph{et al.\ }\cite{taira2003magnetic} suggest that both Ru and Er moments order in a \emph{collinear} AFM structure. At closer inspection it is a linear combination of $\Psi_4$, $\Psi_{11}$ and $\Psi_{12}$. In fact, as these are also powder-diffraction measurements, we assume that the experimental data cannot unambiguously determine the exact linear combination of $\Psi_{11}$ and $\Psi_{12}$. Instead, the situation seems to be akin to Gd$_2$Ru$_2$O$_7$, but with an angle of $\theta_{\mathrm{Ru}}=\theta_{\mathrm{Er}}=54.74^{\circ}$.
As Siddharthan \emph{et al.\ }\cite{siddharthan1999ising} pointed out for rare-earth titanates, the CEF associated with the $D_{3d}$-site symmetry introduces a strong easy-axis anisotropy along the $\hat{z}$ axis for Ho, but less so for Er. This might cause Er$_2$Ru$_2$O$_7$ to attain a AFM 32-pole structure with an easy-$\hat{x}\hat{y}$-plane AFM configuration rather than a spin-ice-like state, despite its perhaps comparable strength of dipole-dipole interaction as in Ho$_2$Ru$_2$O$_7$. 

For the Ru sublattice, $T_N\approx90\,$K \cite{taira2003magnetic} and the size of the Ru moments is reported as $2.(2) \mub$/Ru \cite{taira2002magnetic, taira2003magnetic}. Moreover, the Er moments are reported to order at $T_N\approx10\,$K \cite{taira2003magnetic,gardner2010magnetic}, or $5.4\,$K \cite{taira2002magnetic} and their size is reported to be $4.5 \mub$/Er \cite{taira2003magnetic} at $3\,$K. This is small compared to the value expected in the ionic limit. This reduction has also been observed in other related Er compounds, e.g., Er$_2$Sn$_2$O$_7$\cite{matsuhira2002low} and Er$_2$Ti$_2$O$_7$ \cite{bramwell2000bulk}.   

In our CMP+SDFT calculations, we obtain a degenerate magnetic ground state between a spin-ice-like magnetic structure and a magnetic 32 pole. The former emerges either as combination of $\Psi_1$ and $\Psi_5$, or $\Psi_3$ and $\Psi_7$, while the latter corresponds to any linear combination of $\Psi_{11}$ and $\Psi_{12}$ as shown in \Cref{fig:3} (c) and (b), respectively. The BCs in braces are much weaker, and seem to be a result of $f$-$d$ exchange interaction. We see that the size of the Er moments is overestimated compared to the experimental value, and that the Er moments tend towards the spin-ice-like configuration. Therefore, we expect that if SDFT would not overestimate the size of the Er moment, the 32-pole order that the Ru moments prefer would prevail.

Below $10\,$meV, we only find (meta-)stable magnetic structures that have a finite net magnetization. However, when we constrain our analysis to AFM structures, the 32 pole with $\Psi'_{11}$ and $\Psi_{11}$ is indeed the most stable. Curiously, the two most stable AFM structures are both constructed by $\Psi'_{11}$ and $\Psi_{11}$, but yield a relatively large energy difference of $>10\,$meV owing to the relative orientation of BCs to each other. In particular, parallel alignment with a $\Psi'_{11} \otimes \Psi_{11}$ state has a lower total energy than the $\Psi'_{11} \otimes (-\Psi_{11})$ state. The difference in energy $\Delta E_\mathrm{tot}>10\,$meV is a direct measure of the $f$-$d$ exchange energy.

At about $70\,$meV, we finally obtain a linear combination of $\Psi_4$ and $\Psi_{12}$, which is consistent with the experimental observation up to the angle $\theta$. Experimentally $\theta\approx54.74^{\circ}$, but here we find  $\theta_{\mathrm{Er}}\approx68^{\circ}$ for Er and $\theta_{\mathrm{Ru}}\approx85^{\circ}$ for Ru. Additionally, we note that the Ru moment predicted by SDFT is smaller than the one observed experimentally, $1.3\mub$/Ru $< 2.(2) \mub$/Ru \cite{taira2002magnetic, taira2003magnetic}. As a side remark, we mention here that for other isostructural compounds, e.g., Gd$_2$Ru$_2$O$_7$ \cite{taira2002magnetic,yao2011hydrothermal}, Tb$_2$Ru$_2$O$_7$ \cite{taira2002magnetic,yao2011hydrothermal,xiao2010berry},  and Ho$_2$Ru$_2$O$_7$ \cite{taira2002magnetic,wiebe2004magnetic,yao2011hydrothermal} the reported size of the Ru moments vary depending on the authors.
Further, investigation on different time scales would thus be beneficial.

\subsection{Summary}

This closes the discussion about the magnetic structure of both classes of ruthenates, $A_2$Ru$_2$O$_7$ and $R_2$Ru$_2$O$_7$. To summarise, our results show that (i) $A_2$Ru$_2$O$_7$ with $A=$Ca and Cd prefers the AIAO structure, (ii) $R_2$Ru$_2$O$_7$ with $R=$Pr, Gd, and Er feature an easy-plane 32-pole AFM order with varying angle $\theta$ due to a finite contribution from the AIAO structure, (iii) the magnetic ground state of Ho$_2$Ru$_2$O$_7$ is spin-ice-like, and (iv) compared to the experimental value within SDFT the size of the magnetic dipole moments per site is (a) well-estimated for Ru and Ho in Ho$_2$Ru$_2$O$_7$, (b) overestimated for Ru in $A_2$Ru$_2$O$_7$, as well as for Pr and Er, and (c) underestimated for Ru in $R_2$Ru$_2$O$_7$ with $R=$Pr and Er, while for the Gd$_2$Ru$_2$O$_7$ the experimental data to make a direct comparison has not been reported. Albeit the main focus in the next section is on electronic properties, the size of the moments is also briefly discussed.

\section{Electronic properties} \label{sec:Electronic properties}

From experimental observations, it is clear that electronic correlations play an important role in $A_2$Ru$_2$O$_7$ and $R_2$Ru$_2$O$_7$. Particularly, 
the resistivity
shows metallic behavior for Ca$_2$Ru$_2$O$_7$ and insulating behavior for Pr$_2$Ru$_2$O$_7$ \cite{kaneko2020,kaneko2021fully}. 
This tendency can be confirmed by means of the optical conductivity, which shows a gap for Pr$_2$Ru$_2$O$_7$ and no gap for  Ca$_2$Ru$_2$O$_7$ up to experimental uncertainty. 
Hence, despite the success in reproducing the experimental magnetic structure based on CMP+SDFT in the previous section, above observations call for a careful treatment of electronic correlations beyond SDFT. 

Here, we focus on the Ru-t$_{2\mathrm{g}}$ manifold that forms the conduction bands in SDFT and governs the electronic properties of $A_2$Ru$_2$O$_7$ and $R_2$Ru$_2$O$_7$. Among the discussed compounds, we pick two representative ones: Ca$_2$Ru$_2$O$_7$ and Pr$_2$Ru$_2$O$_7$. The main motivation is given by recent works by Kaneko {\it et al.} \cite{kaneko2020,kaneko2021fully}, where (Ca$_{1-x}$Pr$_x)_2$Ru$_2$O$_7$ has been investigated and the doping parameter $x$ could be continuously tuned. Yet, let us emphasize at this point that Ca$_2$Ru$_2$O$_7$ and Cd$_2$Ru$_2$O$_7$ have demonstrated a great similarity in the SDFT-total-energy landscape discussed in the previous section. Furthermore, Pr$_2$Ru$_2$O$_7$ is the most natural choice among the rare-earth ruthenates, when focusing on the Ru-t$_{2\mathrm{g}}$ bands as we do here. That is because Pr moments do not order and it is thus well-justified to neglect strong $d$-$f$ exchange interaction when considering the electronic properties.

Finally, before constructing a low-energy effective model and solving it within DMFT, let us recall that the central difference between Ca$_2$Ru$_2$O$_7$ and Pr$_2$Ru$_2$O$_7$ is that the Ru-t$_{2\mathrm{g}}$ bands are occupied by 3 and 4 electrons, respectively. It is noteworthy, that experimentally the Ru-t$_{2\mathrm{g}}$ bands in Pr$_2$Ru$_2$O$_7$ are more strongly correlated than the half-filled case of Ca$_2$Ru$_2$O$_7$. This is unexpected for correlated, multiorbital systems \cite{georges2013strong}. The underlying mechanisms that introduces this counterintuitive tendency are elucidated in the following.

\subsection{Low-energy effective models for \texorpdfstring{Ca$_2$Ru$_2$O$_7$}{Ca2Ru2O7} and \texorpdfstring{Pr$_2$Ru$_2$O$_7$}{Pr2Ru2O7}}

To derive the low-energy effective models for the Ru-t$_{2\mathrm{g}}$ manifold for Ca$_2$Ru$_2$O$_7$ and Pr$_2$Ru$_2$O$_7$, we employ a combination of the maximally-localized Wannier functions~\cite{marzari1997maximally,souza2001maximally} and the cRPA method~\cite{aryasetiawan2004frequency}. 
The construction of Wannier functions and the cRPA calculations are performed with an open-source package RESPACK~\cite{nakamura2021RESPACK,RESPACK_URL}. 
In the cRPA calculations, RESPACK employs the band disentanglement scheme proposed in Ref.~\cite{sasioglu2011effective}. 

\begin{figure}
  \includegraphics[width=\columnwidth]{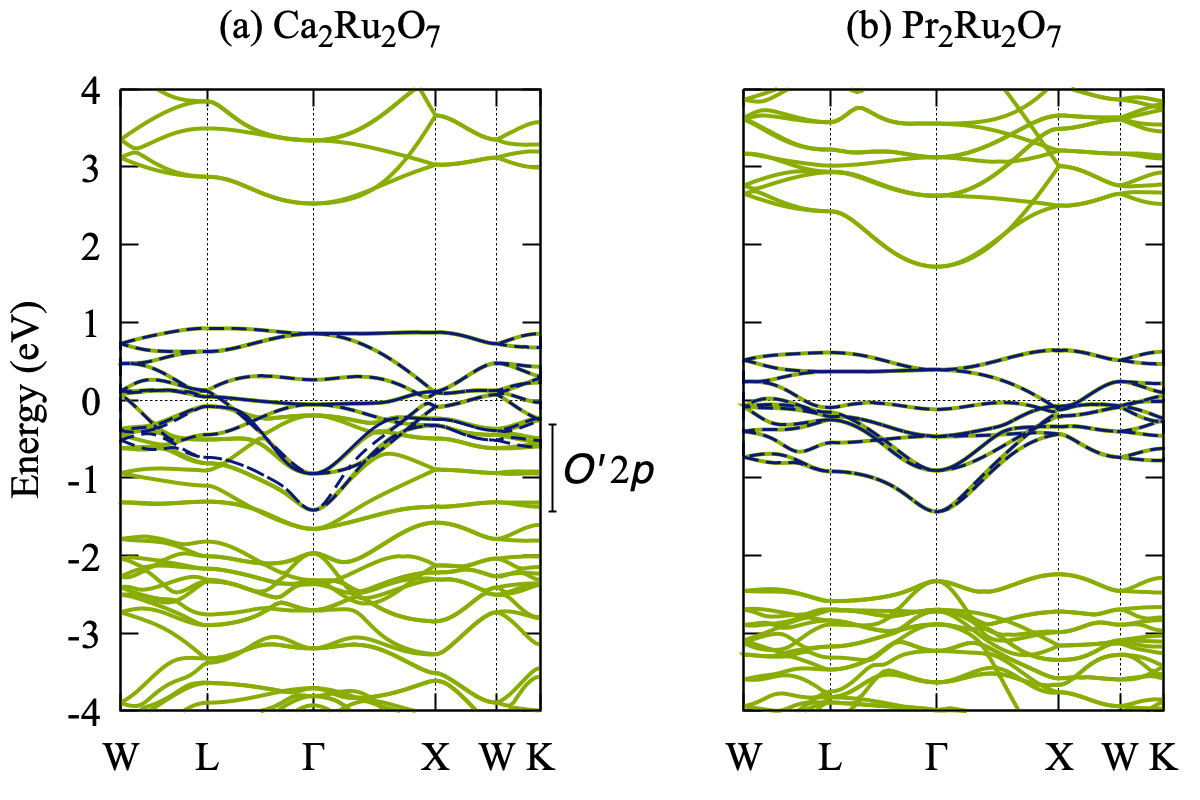}
  \caption{Band structure for (a) Ca$_2$Ru$_2$O$_7$ and (b) Pr$_2$Ru$_2$O$_7$. The green lines are based on a DFT calculation, while the blue dashed lines are based on maximally-localized Wannier functions for the Ru-t$_{2\mathrm{g}}$ orbitals. For Ca$_2$Ru$_2$O$_7$, the entangled O'-$2p$ bands, c.f., \Cref{tab:1}, introduce additional screening.
  \label{fig:4}
  }
\end{figure}

We first perform the DFT-band-structure calculations for Ca$_2$Ru$_2$O$_7$ and Pr$_2$Ru$_2$O$_7$ using {\textsc{Quantum ESPRESSO}}~\cite{giannozzi2017advanced}~\footnote{\red{The reason to use a different \emph{ab-initio} code here is that RESPACK provides a convenient interface to {\textsc{Quantum ESPRESSO}}. The DFT results between VASP and {\textsc{Quantum ESPRESSO}} show no discernible difference, which serves as an additional cross-validation for our numerical results.}}. The result is shown in~\Cref{fig:4} as green lines. 
The structures for Ca$_2$Ru$_2$O$_7$ and Pr$_2$Ru$_2$O$_7$ are taken from Refs.~\cite{munenaka2006novel} and \cite{yamamoto1994crystal}, respectively. 
The optimized norm-preserving Vanderbilt pseudopotentials~\cite{hamann2013optimized} with the PBE (Perdew-Burke-Ernzerhof) exchange-correlation functional~\cite{perdew1996generalized} are downloaded from PseudoDojo~\cite{setten2018pseudodojo} and employed in the DFT calculations.
We use 9$\times$9$\times$9 $\bf{k}$ mesh, and the energy cutoff is set to $100\,$Ry for the wavefunction and $400\,$Ry for the electron-charge density.

The maximally localized Wannier functions~\cite{marzari1997maximally,souza2001maximally} are constructed with the projections of Ru-a$_{1\mathrm{g}}$ and e$'_{\mathrm{g}}$ orbitals. The resulting bands are shown in \Cref{fig:4} as dashed blue lines around the Fermi energy set to $0\,$eV. 
For Ca$_2$Ru$_2$O$_7$, the Ru-t$_{2\mathrm{g}}$ bands are overlapping with the O$'$-$2p$ bands, where O$'$ refers to the oxygen atom at Wyckoff position $8b$ as defined in \Cref{tab:1} and illustrated \Cref{fig:1} (a) and (c). That is why we employ both the outer and inner windows whose energy ranges are $[-1.45\,$eV $:1.5\,$eV$]$ and $[-0.18\,$eV $:1.5\,$eV$]$, respectively. 
For Pr$_2$Ru$_2$O$_7$, the Ru-t$_{2\mathrm{g}}$ bands are isolated from the other bands, and the Wannier functions are constructed from the energy range of $[-1.5\,$eV $:1.5$eV$]$.  
For the constructed Wannier orbitals, we calculate the effective interaction parameters using the cRPA method. 
The polarization function is calculated using $200$ bands with the energy cutoff of $10\,$Ry. 

\begin{table*}
    \caption{cRPA results for the Hubbard interaction $U_{\mathrm{cRPA}}$ for Ca$_2$Ru$_2$O$_7$ and Pr$_2$Ru$_2$O$_7$. To see the effect of the screening, we also show the results of bare interaction $U_{\mathrm{bare}}$. 
    \label{tab:5}
    }
    \begin{indented}
    \item[]\begin{tabular}{@{}lllll}
    \br
                        &                 & $U_{\rm bare}$ (eV)  & $U_{\rm cRPA}$ (eV)  & $U_{\rm cRPA}$/$U_{\rm bare}$ \\ \mr
  Ca$_2$Ru$_2$O$_7$     & a$_{1\text{g}}$ & $9.42$       & $1.10$        & $0.12$    \\ 
                        & e$'_{\text{g}}$ & $9.82$       & $1.16$        & $0.12$    \\ \mr
  Pr$_2$Ru$_2$O$_7$     & a$_{1\text{g}}$ & $11.33$      & $2.36$        & $0.21$    \\  
                        & e$'_{\text{g}}$ & $11.48$      & $2.42$        & $0.21$    \\ \br
    \end{tabular}
    \end{indented}
\end{table*}

The derived Hubbard interaction parameters are listed in~\Cref{tab:5}.
Interestingly, we see a drastic change in $U$ parameters between Ca$_2$Ru$_2$O$_7$ and Pr$_2$Ru$_2$O$_7$: 
Whereas the on-site $U$ is comparable to the bandwidth in the case of Pr$_2$Ru$_2$O$_7$, in the case of Ca$_2$Ru$_2$O$_7$, the $U$ value is much smaller. 
This gives rise to a significant difference in the electronic correlation as discussed below in \Cref{sec:DMFT}. 

The large difference in the $U$ value can be ascribed to the spatial spread of the Wannier orbitals and the strength of the screening. 
In Ca$_2$Ru$_2$O$_7$, the nominal valence of the Ru cations is $5+$, and the energy of Ru-t$_{2\mathrm{g}}$ orbitals is lowered compared to the Pr compound with Ru$^{4+}$ cations due to the stronger attractive potential from the nuclei. 
This draws the energy levels of the Ru-t$_{2\mathrm{g}}$ and O-$2p$ orbitals closer together.
As a result, the hybridization between Ru-t$_{2\mathrm{g}}$ and O-$2p$ orbitals becomes larger, and the Wannier functions become more delocalized in space. 
The difference of the spatial spread of the Wannier orbitals is reflected in the bare $U$ value ($U_{\mathrm{bare}}$) listed in \Cref{tab:5}: We indeed see that the bare $U$ values for Ca$_2$Ru$_2$O$_7$ are smaller than those of Pr$_2$Ru$_2$O$_7$. 

The difference in the screening results in an even more considerable difference in the $U$ values than the difference of the spatial spread of the Wannier orbitals. 
As we see in \Cref{tab:5}, in Ca$_2$Ru$_2$O$_7$, the screening effect is much stronger than that of Pr$_2$Ru$_2$O$_7$, and $U_{\mathrm{cRPA}}$ is about 10 times smaller than $U_{\mathrm{bare}}$. 
This is because the energy levels of O-$2p$ and O$'$-$2p$ bands move closer to the Fermi level. 
In particular, the O$'$-$2p$ bands overlap with the Ru-t$_{2\mathrm{g}}$ manifold in Ca$_2$Ru$_2$O$_7$ and lie very close to the Fermi level, which yields a large contribution to the screening. That is due to the smaller energy that appears in the denominator of expression for the polarization function.

Because the strength of the screening is sensitive to the energy level of the O$'$ bands in Ca$_2$Ru$_2$O$_7$, we may be able to change the $U$ value significantly by controlling the position of the O$'$-$2p$ bands by, e.g., applying external pressure. 
Motivated by this expectation, we perform the cRPA calculations for Ca$_2$Ru$_2$O$_7$ with different lattice constants (internal coordinates are fixed). 
When the lattice constant is $4 \%$ ($2 \%$) smaller, the $U_{\mathrm{cRPA}}$ values become $1.41\,(1.24)\,$eV and $1.45$ $(1.28)\,$eV for Ru-a$_{1\mathrm{g}}$ and Ru-e$'_{\mathrm{g}}$ orbitals, respectively. 
We observe an increase as large as about several tens percent in the value of $U$ by external pressure. 
The result is of great interest because, in Ca$_2$Ru$_2$O$_7$, applying pressure might make the compound more strongly correlated as a result of the drastic increase in $U$, which thus can get larger than the increase in the bandwidth. 
This is in stark contrast with a common belief that the external pressure weakens the electronic correlation. That is based on the fact that in usual materials, the bandwidth increases, whereas the $U$ value gets less affected. 
Such an interesting effect would also be seen in other Ru$^{5+}$ compounds, i.e., A$_2$Ru$_2$O$_7$, such as Cd$_2$Ru$_2$O$_7$. 
Indeed, there is an experimental report that applying chemical or external pressure unexpectedly induces a MIT in the Cd compound~\cite{jiao2018effect}.

\subsection{DMFT results}
\label{sec:DMFT}

Based on the maximally-localized Wannier functions and cRPA results discussed in the previous subsection, we construct low-energy effective models for Ca$_2$Ru$_2$O$_7$ and Pr$_2$Ru$_2$O$_7$, and solve them within the DMFT.
The DMFT impurity problem is solved employing the finite-temperature extension \cite{capone2007,liebsch2012} of the exact diagonalization method \cite{caffarel1994}, where the dynamical mean field is represented by 9 bath sites.

\begin{table*}
    \caption{Orbital occupancy at a Ru site in the AIAO state calculated within DMFT at $T\approx50\,$K. Note that the e$'_{\mathrm{g}}$ orbitals are doubly degenerate.
    \label{tab:dmft}
    }
    \begin{indented}
    \item[]
    \begin{tabular}{@{}lllll}
    \br
                        &       orbital          & majority spin  &  minority spin & total \\ \mr
  Ca$_2$Ru$_2$O$_7$     & Ru-a$_{1\mathrm{g}}$ & $0.78$      & $0.36$        & $1.14$    \\ 
                        & Ru-e$'_{\mathrm{g}}$ & $1.25$       & $0.60$       & $1.85$    \\ \mr
  Pr$_2$Ru$_2$O$_7$     & Ru-a$_{1\mathrm{g}}$ & $1.00$      & $0.99$        & $1.99$    \\  
                        & Ru-e$'_{\mathrm{g}}$ & $1.94$      & $0.08$        & $2.01$    \\ \br
    \end{tabular}
    \end{indented}
\end{table*}

For Ca$_2$Ru$_2$O$_7$, although the zero-temperature magnetic structure is controversial in experiments, our CMP+SDFT results suggest that the AIAO structure shown in \Cref{fig:2} is the magnetic ground state. 
Therefore, we assume that the AFM exchange interaction between Ru sites is of the AIAO type in the DMFT calculation.
This allows to copy the self-energy matrix with respect to spin and orbital degrees of freedom, that is obtained by solving the impurity problem at one Ru site, to the other three Ru sites in the crystallographic unit cell by taking into account the appropriate rotation in the spin-orbital space. Generally, this setup permits to converge to either a spin-polarized AFM AIAO state or non-spin-polarized state.

In the low temperature regime at $T\approx50\,$K, we indeed converge to the AFM AIAO state, where both Ru-a$_{1\mathrm{g}}$ and Ru-e$'_{\mathrm{g}}$ orbitals are nearly half-filled with an occupancy of approximately 1 and 2 electrons, respectively. The exact values are presented in \Cref{tab:dmft}. Obviously, the occupancies sum up to 3 electrons within the Ru-t$_{2\mathrm{g}}$ manifold. 
At each Ru site, there is a majority spin and a minority spin, which refers to the local spin-up and spin-down state according to the AIAO structure. From \Cref{tab:dmft}, we see that each orbital at a Ru site is not fully spin-polarized but yields an on-site magnetic moment of $1.07\mub$/Ru, which will be discussed below.   

\begin{figure}
  \includegraphics[width=\columnwidth]{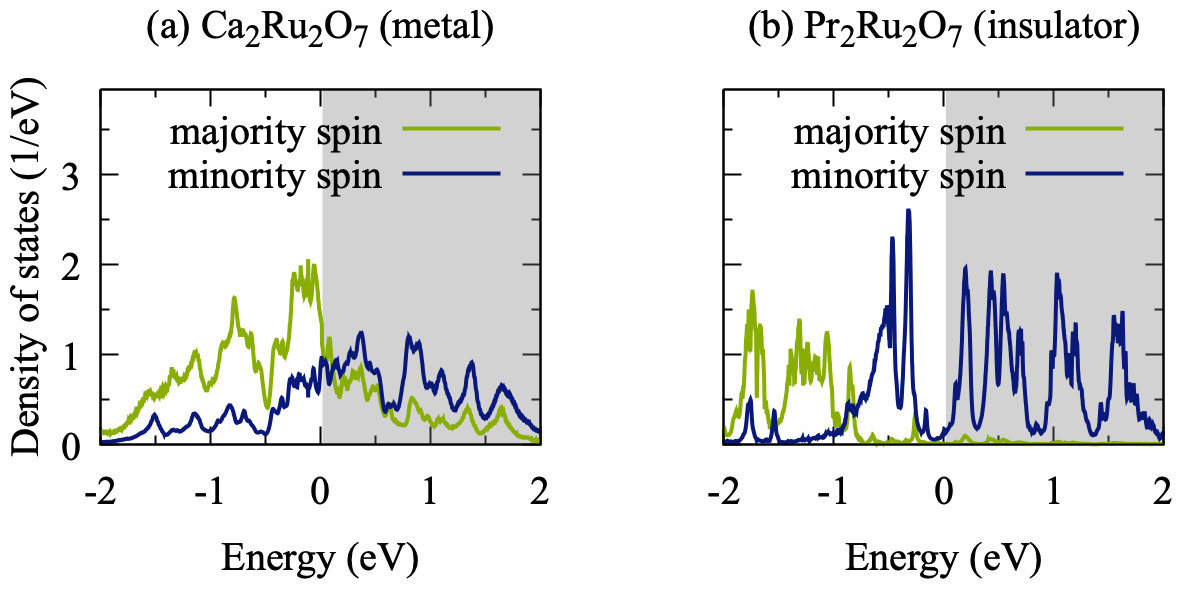}
  \caption{Local density of states at one Ru site, calculated with the DMFT for the low-energy effective models for (a)~Ca$_2$Ru$_2$O$_7$, and (b)~Pr$_2$Ru$_2$O$_7$.}
  \label{fig:5}
\end{figure}

\Cref{fig:5} (a) shows the local density of states at the Ru site in Ca$_2$Ru$_2$O$_7$. The accumulation of majority-spin states below the Fermi energy at $0\,$eV yields a large spin-polarization. Additionally, we see that no gap appears at the Fermi level and the DMFT solution is indeed metallic. This is due to the relatively small value of $U$, which is a consequence of the substantial screening by the O$'$-$2p$ bands. Overall, this is consistent with the observed bad-metallic behavior in the experimentally measured resistivity \cite{kaneko2021fully}.

On the other hand, following the same procedure for Pr$_2$Ru$_2$O$_7$ yields an insulating DMFT solution as seen in \Cref{fig:5} (b). Note that also here we assume the AFM AIAO state, which is justified by the fact that the insulating behavior is expected to be independent of the choice of the magnetic structure as it appears in all compounds of $R_2$Ru$_2$O$_7$. Additionally, we see in SDFT that the AIAO state is metastable so that we can obtain a converged solution. Nevertheless, as discussed in \Cref{sec:Magnetic structure} the ground-state magnetic order is a 32-pole with small contributions from the AIAO structure. By means of our DMFT results, the size of the band gap can be estimated to be about $0.2\,$eV. This agrees well with the reported value based on the optical-conductivity measurements \cite{kaneko2020}, i.e., $0.25\,$eV. 
The formation of the band gap in Pr$_2$Ru$_2$O$_7$ is ascribed to the large value of the screened interactions $U_{\mathrm{cRPA}}$ compared to the more substantially screened, and thus smaller, value of $U_{\mathrm{cRPA}}$ in Ca$_2$Ru$_2$O$_7$, c.f., \Cref{tab:5}. Thus, the strong electronic correlation in Pr$_2$Ru$_2$O$_7$ drives it into an insulating state, while the stronger screening in Ca$_2$Ru$_2$O$_7$ allows the metallic state to prevail.

From \Cref{tab:dmft}, one can see that the Ru-a$_{1\mathrm{g}}$ orbital is basically fully occupied in Pr$_2$Ru$_2$O$_7$, while the Ru-e$'_{\mathrm{g}}$ orbitals are half-filled. Therefore, the spin-polarization emerges on the latter, in contrast to the case of Ca$_2$Ru$_2$O$_7$ where all Ru-t$_{2\mathrm{g}}$ bands contribute. Let us recall that, for Ca$_2$Ru$_2$O$_7$ the effective Ru moment is reported as $0.60\mub$/Ru based on $\mu$SR results \cite{miyazaki2010magnetic}, where also glasslike randomness was detected. The Ru moment defined in SDFT is $1.17\mub$/Ru, as discussed in \Cref{subsec:Magnetic structure A}, which is clearly larger than the experimental value. Within DMFT we obtain $1.07\mub$/Ru for the Ru moment. That is a minor reduction and is still larger than the experimental value. For Pr$_2$Ru$_2$O$_7$, the the Ru moment is (i) 1.48(4)$\mu_B$/Ru in experiment \cite{van2017induced}, (ii) 0.94$\mu_B$/Ru in SDFT, and (iii) 1.87$\mu_B$/Ru in DMFT.
In both cases, we see that the inclusion of local quantum fluctuations within DMFT improves upon the predicted magnetic moment although the deviation from the experimental value cannot be fully amended. That might point towards more intriguing mechanisms based on spin-dynamic effects and magnetic frustration that are not yet taken into account on the level of the present low-energy effective model solved within DMFT.

\section{Conclusion} \label{sec:Conclusion}

We have described the magnetic structure and electronic properties of $R_2$Ru$_2$O$_7$ with $R^{3+}=$ Pr, Gd, Ho, and Er, as well as $A_2$Ru$_2$O$_7$ with nonmagnetic $A^{2+}=$ Ca, and Cd from first-principles. 
That is, with Er$_2$Ru$_2$O$_7$, we tackled the most challenging compound that was included in the data set of the high-throughput calculation in which some of the present authors introduced the CMP+SDFT prediction scheme \cite{huebsch2021benchmark}. Initially, it seemed CMP+SDFT fails to treat Er$_2$Ru$_2$O$_7$ well, and by extension we expected that $R_2$Ru$_2$O$_7$ is generally not well-described within this scheme. Motivated by the interest of Kaneko {\it et al.} \cite{kaneko2020,kaneko2021fully} in (Ca$_{1-x}$Pr$_x)_2$Ru$_2$O$_7$, we once again faced this class of materials. To our great surprise and delight, we found good agreement of our numeric results and the available experimental data at closer inspection.

For the magnetic ground state, we have demonstrated that the discussion of magnetic structures featured in cubic-pyrochlore ruthenates greatly benefits from the classification in terms of the CMP theory. Firstly, the easy-plane AFM structure realized in $R_2$Ru$_2$O$_7$ with $R=$Pr, Gd, and Er corresponds to a 32 pole. Due to the easy plane and ambiguity in the powder-diffraction measurements, there is a particular need to take care when comparing the numerical and experimental results, which is done within the scope of this work. Secondly, the spin-ice structure is a linear combination of a dipole and octupole that both have T$_{1\mathrm{g}}$ symmetry. 
And lastly, the AIAO is prominently classified as the A$_{2\mathrm{g}}$ octupole. The CMP+SDFT scheme successfully leads to the identification of key tendencies when varying $A$ and $R$ elements. This is most significantly shown by the fact that, the magnetic ground state of Ho$_2$Ru$_2$O$_7$ is spin-ice-like, while the other rare-earth ruthenates---$R_2$Ru$_2$O$_7$ with $R=$Pr, Gd, and Er---feature an easy-plane 32-pole AFM order. Also, our data shows that $A_2$Ru$_2$O$_7$ with $A=$Ca and Cd prefers the AIAO structure.

For the electronic properties, band-structure and cRPA calculations reveal that the relative energy of O-$2p$ and O$'$-$2p$ bands with respect to the Ru-\ttwog bands controls the strength of electronic correlation. Building upon that, we constructed a low-energy effective model that can qualitatively and quantitatively account for the metallic state in Ca$_2$Ru$_2$O$_7$ and the Mott-insulating state in Pr$_2$Ru$_2$O$_7$. In particular, the band gap in Pr$_2$Ru$_2$O$_7$ is in good agreement with the optical-conductivity measurement.

Regarding the size of the on-site magnetic moment, regrettably,
it cannot be reliably reproduced. There are varying tendencies towards either over- or underestimation within SDFT depending on the magnetic site. The DMFT calculations based on our low-energy effective model for Ca$_2$Ru$_2$O$_7$ and Pr$_2$Ru$_2$O$_7$ slightly improve the agreement of the experimental value compared to the SDFT result. As this does not fully correct the numerical values, we suspect that spin-dynamic effects and magnetic frustration that are beyond the scope here actually play a significant role.

Looking forward, there are still plenty of intriguing mysteries to be solved for pyrochlore ruthenates:
Firstly, it could be interesting to investigate the spin dynamics of these systems. This is not only promising with respect to the predicted size of the on-site magnetic moment, but also to elucidate the experimental evidence of spin glass behavior in Ca$_2$Ru$_2$O$_7$ \cite{taniguchi2009spin, miyazaki2010magnetic} and Cd$_2$Ru$_2$O$_7$ \cite{miyazaki2010magnetic}. Next, it will be interesting to fully account for doping in our calculations in order to catch up with the experimental capability. In this context, there are new magnetic states, complex doping-dependency of the resistivity and a MIT to explore. For instance, (Ca$_{1-x}$Pr$_x)_2$Ru$_2$O$_7$ with small $x$ seems to have a FM ground state. Then, the magnetic order in different temperature regimes deserves a closer look. In fact, for Ho$_2$Ru$_2$O$_7$ the magnetic order on the Ru site shifts
from a spin-ice-like structure to an almost collinear FM at low temperatures where the Ho site orders as a spin-ice-like structure with a small FM canting that aligns AFM with the Ru sublattice. To our knowledge, the experimental high-temperature magnetic structure of other cubic-pyrochlore ruthenates has not been reported and may reveal further surprising shifts. And lastly, it will be interesting to investigate in more detail the relationship between our findings with regards to the O$'$-$2p$ bands and the experimentally observed pressure-induced MIT in Cd$_2$Ru$_2$O$_7$.

\ack
We thank Yoshinori Tokura, Kentaro Ueda, and Ryoma Kaneko for directing our attention to these materials and for valuable discussions.
This work was supported by JSPS KAKENHI (Grant No. JP19H02594 19H05825, 21H04437, 21H04990, 20K14423, and 21H01041) and “Program for Promoting Researches on the Supercomputer Fugaku” (Project ID: hp210163) from  MEXT.

\section*{References}

\bibliography{ruthenate}

\end{document}